\documentclass[12pt]{spieman}  
 \usepackage{amsmath,amsfonts,amssymb}
\usepackage{graphicx}
\usepackage{setspace}
\usepackage{tocloft}
\usepackage{lineno}
\usepackage[noabbrev]{cleveref}
 
\usepackage[colorlinks=true, allcolors=blue]{hyperref}
\usepackage{siunitx}
\usepackage{subcaption}
\usepackage{bm}
\usepackage{cprotect}  
\usepackage{algorithm}
\usepackage{algpseudocode}

\usepackage[dvipsnames]{xcolor}
\usepackage{fancyvrb}
\RecustomVerbatimCommand{\VerbatimInput}{VerbatimInput}%
{fontsize=\footnotesize,
 frame=lines,  
 framesep=2em, 
 rulecolor=\color{Gray},
 label=\fbox{\color{Black}p\_custom\_XXXX.txt},
 labelposition=topline,
 commandchars=\|\(\), 
 commentchar=*        
}

\usepackage{listings}
\definecolor{codegreen}{rgb}{0,0.6,0}
\definecolor{codegray}{rgb}{0.5,0.5,0.5}
\definecolor{codepurple}{rgb}{0.58,0,0.82}
\definecolor{backcolour}{rgb}{0.95,0.95,0.92}

\lstdefinestyle{mystyle}{
    backgroundcolor=\color{backcolour},   
    commentstyle=\color{codegreen},
    keywordstyle=\color{magenta},
    numberstyle=\tiny\color{codegray},
    stringstyle=\color{codepurple},
    basicstyle=\ttfamily\footnotesize,
    breakatwhitespace=false,         
    breaklines=true,                 
    captionpos=b,                    
    keepspaces=true,                 
    numbers=left,                    
    numbersep=5pt,                  
    showspaces=false,                
    showstringspaces=false,
    showtabs=false,                  
    tabsize=2
}

\title{Sequential and non-sequential Zemax Dynamic Link Libraries for generating image slicer integral field units}

\author[a]{Ellen Lee}
\affil[a]{University of Hawai'i at Manoa, Institute for Astronomy, 640 N. Aohoku Place, Hilo, HI, 96720}

\cftpagenumbersoff{figure}
\cftpagenumbersoff{table} 
\begin{document} 
\maketitle

\begin{abstract}
In astronomy, image slicer integral field units (IFUs) are often used in integral field spectrographs to simultaneously record spatial and spectral information. The majority of astronomical instruments, including integral field spectrographs, are designed using the Zemax OpticStudio optical design software. Modeling an image slicer IFU in Zemax traditionally requires using many separate configurations, which is slow, cannot accurately model diffraction, and can prevent one from fully describing their instrument within a single file. This paper presents the implementation of sequential and non-sequential Dynamic Link Libraries (DLLs) that efficiently model image slicer IFUs with a known design. The parameters used to manipulate the surfaces are chosen to match fabrication processes. The DLLs identically reproduce natively transformed surfaces in Zemax and have also been used to replicate the design of SPECTRE, a facility spectrograph with a 36-slice image slicer for the NASA Infrared Telescope Facility. The DLLs also work in transmission and can be used in other applications that require modeling a nearly arbitrary grid of surfaces. In the future, this work may facilitate the creation of an optical design tool for IFUs that starts from basic system requirements.
\end{abstract}

\keywords{Ray tracing, Zemax, spectrographs, diffraction, integral field spectroscopy, image slicers, integral field units, hyperspectral imaging}

{\noindent \footnotesize\textbf{*}Ellen Lee, \linkable{ellenlee@hawaii.edu} }


\section{Introduction} \label{sec:intro}
Despite their optical complexity, integral field spectrographs (IFSs) are attractive to astronomers for their ability to create images that are simultaneously spatially and spectrally multiplexed. They are often used to perform spectral mapping or avoid slit losses incurred by the standard slit spectrograph design.\cite{marsset2020twenty} An image slicer is a type of integral field unit that utilizes an array of tilted mirrors to reformat the focal plane into one or more pseudo-slits that can be spectrally dispersed downstream in the optical path. Image slicers are advantageous because they are inherently achromatic, have low thermal background, and yield a higher filling factor on the detector compared to other types of IFUs.\cite{allington1998sampling,2006NewAR..50..244A,hagen2013review}

\subsection{Optical description} \label{sec:intro.optics}

The essential function of an image slicer IFU is to rearrange the focal plane so that it can be dispersed by the spectrograph without spatial confusion on the detector.\cite{weitzel19963d} The image slicer itself is a stack of ``slicer mirrors" (or simply "slices") that is placed at the input focal plane. We can begin by considering an individual slice and the corresponding the portion of the focal plane that illuminates that slice (a ``field channel"). The slicer mirror may have optical power that acts on the pupil rather than the image, reimaging the pupil to the next optic and controlling its position and size. The tilt of the slicer mirror sets the position of the pupil image. Next, a ``pupil mirror" is placed at this pupil image to refocus the light and steer the beam into the desired position. The final element in the IFU is an optional ``field mirror" near the re-imaged slice that may control the position and size of the exit pupil, removing telecentricity errors. As we will discuss momentarily, this simplified explanation does not encapsulate the diversity of real image slicer designs which can have flat slicer and/or pupil mirrors along with additional optical components in some cases.

\begin{figure}
    \centering
    \includegraphics[width=0.8\linewidth]{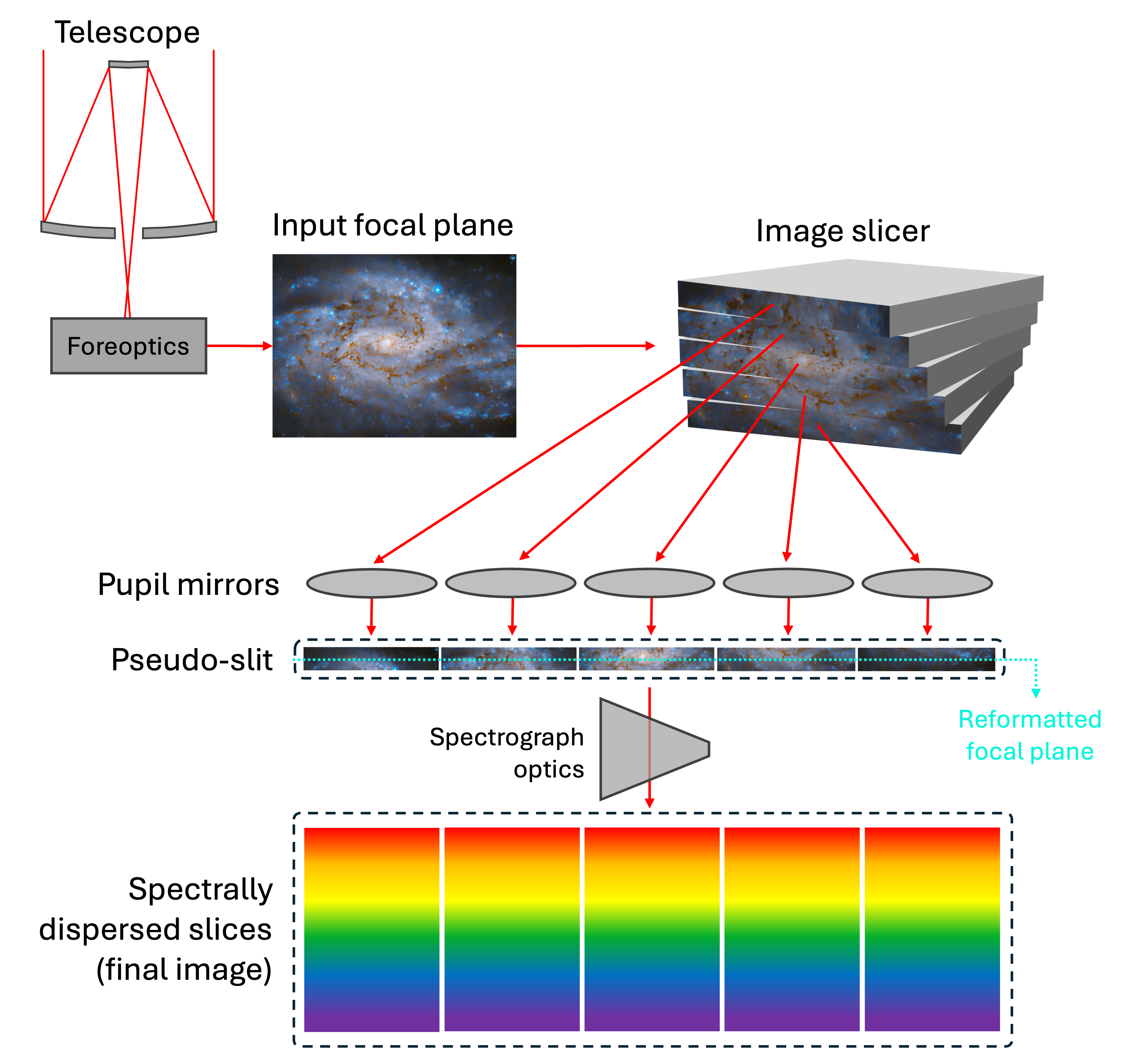}
    \caption{The basic function of an image slicer IFU is to reformat the focal plane into one or more pseudo-slits. The pupil mirrors are shown as transmissive elements and only a single pseudo-slit is shown at the reformatted focal plane for simplicity. The IFS is optically similar to a classical slit spectrograph except with the entire focal plane rearranged by the IFU to go through the slit mask(s). (This diagram takes considerable inspiration from the illustrations in Refs.~\citenum{2006NewAR..50..244A} and \citenum{vives2006set}.)}
    \label{fig:ifs_diagram}
\end{figure}

\begin{figure}[h]
    \centering
    \includegraphics[width=0.75\linewidth]{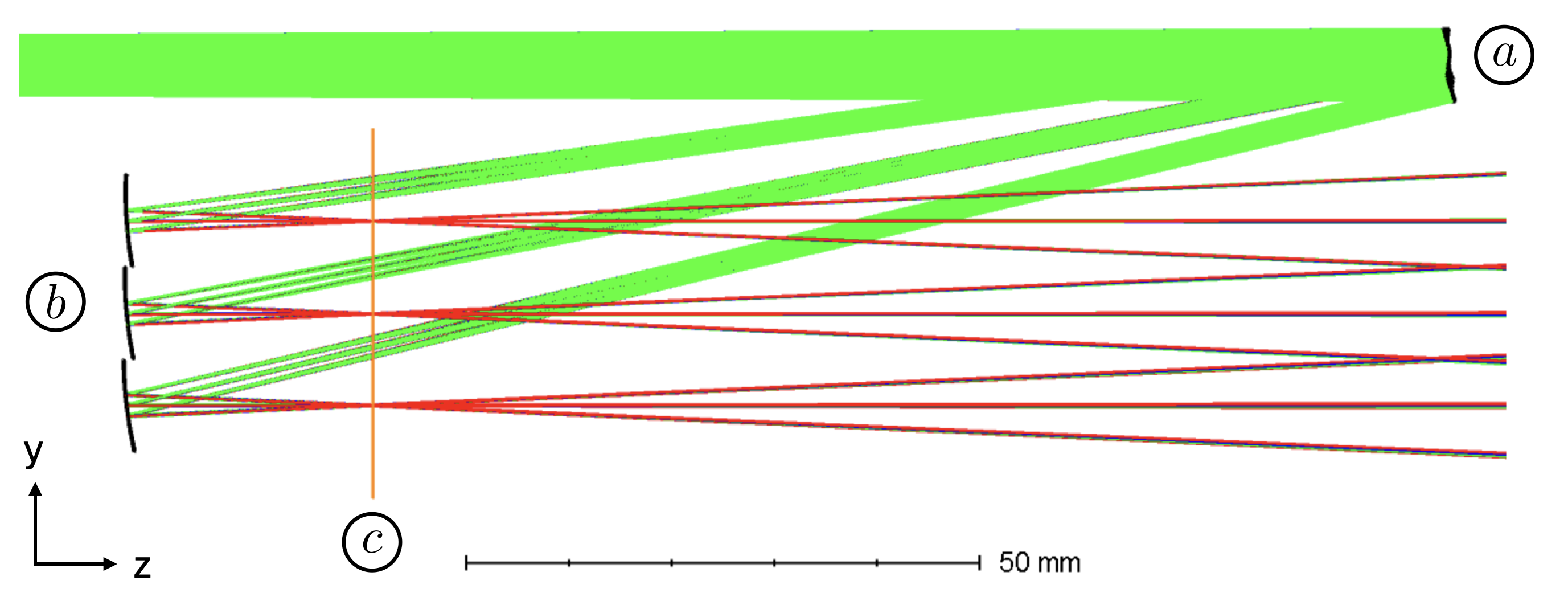}
    \caption{Zemax ray trace of SPECTRE's image slicer IFU. a) The image slicer is placed at the input focal plane to image the pupil for each field channel. b) Pupil mirrors are placed at the resulting pupil images to rearrange and reimage the focal plane. c) The intermediate focal plane contains three rows of pseudo-slits. SPECTRE lacks the field mirrors that would often be located here.}
    \label{fig:ifu-labeled}
\end{figure}

\begin{figure}
    \centering
    \includegraphics[width=0.85\linewidth]{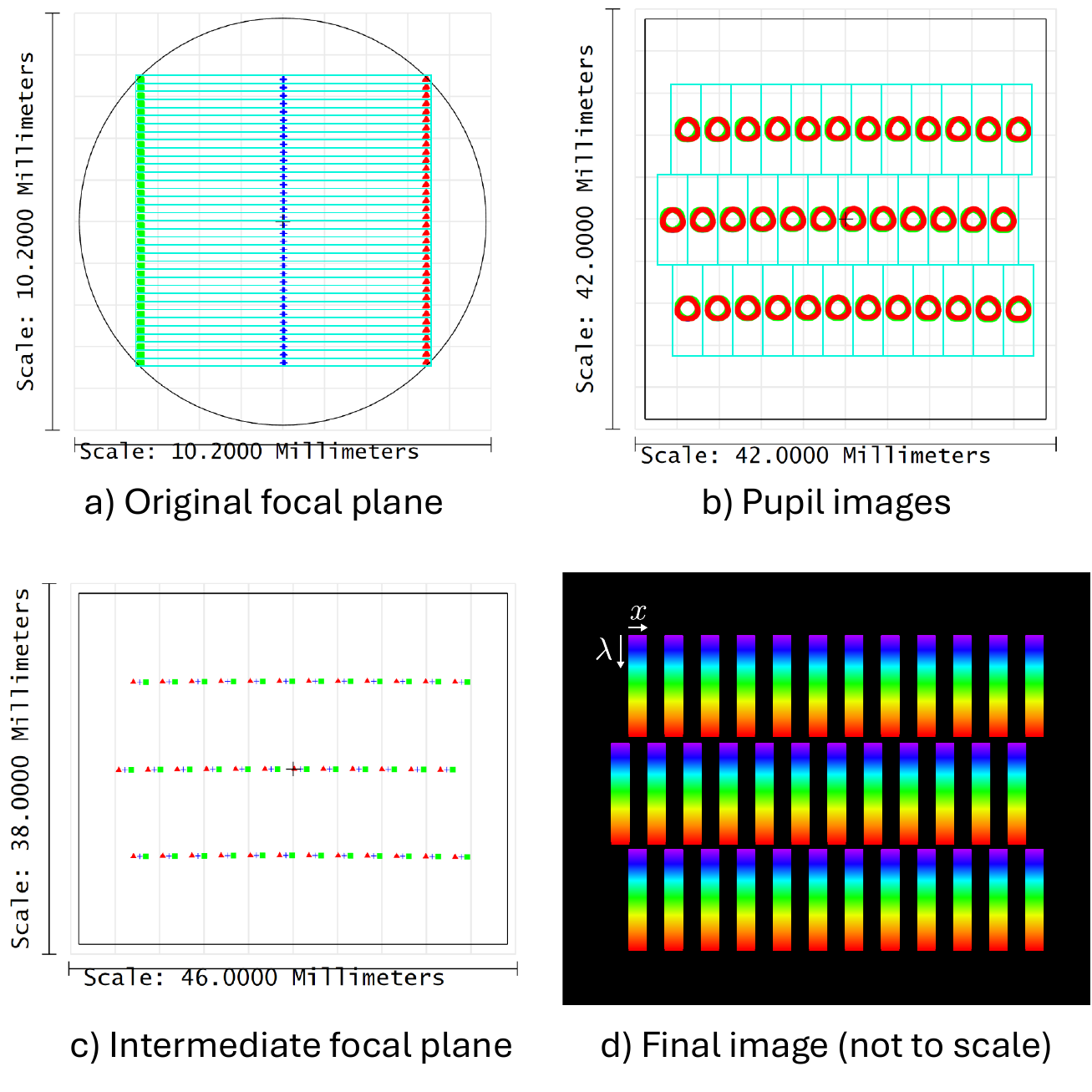}
    \caption{Image and pupil planes in SPECTRE's IFU. a) At the input focal plane, the positions of the 36 slicer mirrors are outlined in cyan. Three points are sampled along each slice. b) Unstacked pupil images. Each of the pupil mirrors is shown in cyan. (The pupil images are circular, but appear slightly deformed due to a drawing defect in Zemax when the footprint diagram is zoomed out.) c) The image has been rearranged into three rows of pseudo-slits. A single wavelength is shown. d) The final image on the spectrograph's detector. Each streak is an image of the slice dispersed in wavelength, akin to what one would see if 36 slit spectrographs were combined.}
    \label{fig:ifu-imgs}
\end{figure}

The arrays of slicer, pupil, and optional field mirrors are tilted in a manner to rearrange the field channels so that they lie side by side in the output focal plane of the IFU, which can be called the ``intermediate" or ``reformatted" focal plane. The result is one or more ``pseudo-slits" that can be dispersed by the spectrograph optics downstream of the IFU. This principle was first described by Ref.~\citenum{content1997new} and is illustrated in Fig.~\ref{fig:ifs_diagram}. As an example, Figs.~\ref{fig:ifu-labeled} and \ref{fig:ifu-imgs} show the IFU inside of SPECTRE, an IFS designed for the NASA Infrared Telescope Facility.\cite{connelley2022spectre} SPECTRE's IFU uses two sets of reflective surfaces--the image slicer and pupil mirrors, with no field mirrors--to create three rows of pseudo-slits.

In practice, each IFS has its own peculiarities such that there is not really a typical image slicer design. Fabrication techniques motivate the layout of image slicers, which are designed to meet image quality requirements across the focal plane while minimizing the cost. Although the slicer mirrors are often powered, they can instead be flat if the optical power is delivered by downstream collimating lenses or mirrors (e.g., DKIST/DL-NIRSP\cite{lin2022misi}). It is also not uncommon for the pupil mirrors to be flat to help control image quality, as in TMT/IRIS \cite{zhang2018infrared} and Keck/KWCI. The IFUs in the older VLT/SPIFFI\cite{tecza2000spiffi} and 3D\cite{weitzel19963d} only use flat surfaces in both the image slicer and pupil mirrors for manufacturing purposes, meaning that there is no properly re-imaged focal plane after the IFU. There are also geometric considerations such as in VLT/MUSE,\cite{henault2004slicing} which uses four identical stacks of slices placed side by side to match the rectangular input image plane.

\subsection{Design challenges} \label{sec:intro.challenges}

Designing an image slicer IFU is challenging for two major reasons. First, the design process is driven by a plethora of optical and mechanical considerations. The field of view, F-ratio, desired spatial sampling, interfaces with the telescope and downstream spectrograph, physical restrictions in the instrument, and fabrication process among other factors influence the design of the IFU. The exact shape and position of each mirror (or lens) element in the IFU can be optimized to meet image quality requirements. Stray light must also be carefully mitigated as the numerous sharp, reflective edges in the IFU can easily generate unwanted light paths. From Sec.~\ref{sec:intro.optics}, it is clear that the unique requirements of different systems result in diverse morphologies for image slicer IFUs. There is currently no standardized process for designing an image slicer IFU from fundamental requirements, nor a common reference for how all potential requirements should be defined and propagated to a manufacturing prescription for the IFU.

Next, even if the design of the IFU is known, it is difficult to represent in optical design software such as Zemax.
IFSs are usually designed with the sequential ray tracing mode of Zemax, where rays are assumed to propagate from one surface to the next in a prepared sequence. 
Although sequential mode comes with powerful optimization tools, an image slicer cannot be represented using native surface types. Sequential mode is therefore limited in its ability to model and optimize image slicers, particularly in the physical optics domain.
In contrast, non-sequential mode allows the user to compose geometrically complex objects like image slicers. It is flexible in that it allows rays to interact with objects in an arbitrary order multiple times, but is typically only used for stray light analysis after the optical system has been designed due to a significant loss of analysis features.

The present work aims to address this second problem of representing an image slicer IFU with a known prescription in Zemax. In other words, the goal is to create a software that takes a pre-defined design for an image slicer IFU and reproduces its optical behavior. This will enable the eventual creation of an optical design tool for image slicers, which would start from basic requirements and generate parameters for the IFU that can be passed to the interface described in this paper.

\subsection{Existing methods for representing image slicers in Zemax} \label{sec:alternatives}
In sequential mode, a common workaround is to design each field channel in a separate ``configuration." However, the resulting ray trace is computationally expensive, which may severely inhibit optimization unless only a few field channels are included in the file. Physical optics features in Zemax can only be applied to single configurations, so the combined diffraction from the entire IFU cannot be analyzed. The use of configurations for the image slicer prevents one from describing other parts of the instrument that may also need separate configurations, such as different spectral channels, because configurations cannot be nested. Spectral channels must be designed in separate files. Finally, the maximum number of configurations that is supported by Zemax limits the number of field channels, which may also necessitate using multiple files if the number of slices is large. Any changes to the remainder of the optical system must be carefully propagated to all of the files. This is tedious at best and can result in design inconsistencies at worst, both of which cost many of hours of the optical designer's time.

It is common to integrate the field channels into a single configuration using ``mixed mode," which allows one to use non-sequential groups within a sequential system. For image slicers that have initially been designed using multiple configurations, one can export every configuration to a CAD file, merge the surfaces in an external program like SolidWorks, and re-import the object as a non-sequential group. This process comes with its own idiosyncrasies. If there are points of intersection between adjacent slice surfaces, it may be difficult to merge the surfaces into a solid object. Ray tracing accuracy and speed can also be degraded depending on how the surface is represented in the CAD file.\cite{zmxcad} The user should also be wary of discrepancies in coordinate systems between the re-imported object and Zemax, since they may have different origins or be flipped along the z-axis.

Rather than re-importing the image slicer as a CAD file, one can also use a series of native non-sequential objects to compose the image slicer as in Ref.~\citenum{liu2022hybrid}. This may be preferable in some aspects because built-in surfaces provide faster and more robust ray traces. However, this process requires the user to manually position the slices or write a script that updates slice parameters using the ZOS-API,\cite{zosapi} which requires significant coding ability.

Several problems arise when using mixed mode. Non-sequential ray tracing is slower than for a purely sequential system. More pertinently, using mixed mode results in the loss of all analysis features that require the paraxial ray trace and removes the ability to simulate diffraction patterns. Furthermore, the design of the image slicer cannot be easily modified once it is converted to a non-sequential component, rendering optimization of the IFU impossible. All are noticeable drawbacks when one is interested in evaluating and optimizing image quality, as is the case for imaging spectroscopy.

\subsection{Implementing a custom surface type for image slicers}
It is possible to represent custom surfaces in Zemax using a DLL, which is a pre-compiled program written in C that describes optical behavior. Zemax allows several types of DLLs, notably user-defined surfaces which allow one to specify a sequential ray trace through an arbitrary surface. User-defined objects are analogous to user-defined surfaces for non-sequential mode. The user can interact with the DLL in the same manner as built-in surface types in Zemax. This includes an interface for updating parameters that define the surface and thus does not require coding ability from the user.\cite{zmxDLL}

A DLL can be used to model an entire image slicer within a single configuration with no practical limits on the number of slices. It also becomes possible to perform optimization and to use physical optics propagation to analyze the combined diffraction of the entire IFU. Additionally, the reduction in number of configurations makes ray tracing significantly faster, which speeds up tolerancing and thermal analysis. Configurations can be preserved for other uses in the spectrograph. Drawbacks include a marginal reduction in ray tracing speed compared to using built-in surfaces and the potential for Zemax to crash if the DLL is not implemented robustly.

Unsurprisingly, a DLL for generating image slicers already exists in the literature. Ref.~\citenum{vives2006set} implements a column of spherical or flat slicer mirrors; it notably appears to have been used in the design of the Keck Planet Finder \cite{gibson2018keck}. As techniques for fabricating image slicers have improved and diversified over the past two decades, more complex designs and layouts have been used in astronomical instruments. The approach presented in this paper is generalized in that the DLLs can describe a nearly arbitrary grid of surfaces that are tilted and translated about different axes. As a result, there is much greater flexibility in the shape and orientation of each slice as well as the layout of the image slicer. Because the rows of the grid can be offset from each other, is possible to generate the downstream arrays of surfaces that are necessary to re-image the focal plane from the multitude of pupil images created by the image slicer.

\subsection{Overview}
In this report, I present the implementation of a set of sequential and non-sequential DLLs that model a grid of flat, rotationally symmetric, or cylindrical conicoid surfaces. The primary goal of this paper is to drastically improve the ray tracing speed for image slicer IFUs and allow them to be fully described in Zemax without the limitations of the existing methods described in Sec.~\ref{sec:alternatives}. Sec.~\ref{sec:overview} provides an overview of how the DLLs are designed. Then, Sec.~\ref{sec:verification} compares the DLLs to identical surfaces produced natively in Zemax and verifies their ability to reproduce the design of an existing IFU instrument. The remainder of this paper describes the implementation of the sequential (Sec.~\ref{sec:implementation}) and non-sequential (Sec.~\ref{sec:nsc}) DLLs in enough detail to reproduce the software. The source code is publicly available on Github. Refer to the Code, Data, and Materials Availability statement at the end of this paper for more information.

\section{Methodology} \label{sec:overview}
We will now discuss image slicers in the specific context of how they are generated by the DLLs. Although the modeled surface will be referred to as an image slicer, the DLLs can be used to express an arbitrary rectangular grid of powered or flat surfaces including the pupil and field mirrors.

\subsection{Fabrication processes as a motivation for parameters}
A challenge of modeling image slicers is that they require a large number of parameters to fully describe. The parameters chosen for this work are motivated by common manufacturing techniques, but the DLLs are generally agnostic to which process is chosen.

We will adopt a piece of non-standard terminology to reflect the fabrication processes of Bertin Technologies. This is because they are manufacturing SPECTRE's IFU, which has been a primary motivation for this paper. Their process for manufacturing image slicers generally involves polishing a stack of slices together as a single surface and then rotating each slice relative to each other to produce the image slicer, akin to a stack of books.\cite{vives2008new} A stack of slices that originate from the same parent surface will be referred to as a ``section" in this paper. Each section produces a row of pupil images. For instance, SPECTRE contains three sections which produce three vertically offset rows of pupil images. The 12 slices in each section are polished together, and then each is rotated by a slightly different amount to unstack the pupil images along the horizontal direction.

Free-form image slicers that are produced by other fabrication processes cannot necessarily be specified in terms of sections. Durham University is known for machining monolithic image slicers for use at infrared wavelengths, which comes at a significantly reduced cost and fabrication time than other techniques.\cite{content1998advanced} For example, they have produced the image slicer for the SCALES infrared spectrograph on the W. M. Keck Observatory \cite{kupke2022scales}. Canon Inc. has also been developing a 5-axis CNC machining technique that can produce freeform optical surfaces.\cite{sukegawa2023ultra} An example is the IFU in Flare Sentinel, which is intended for use in space missions for solar astronomy.\cite{lin2023flare} Such techniques are more permitting in many aspects by allowing for more complex and numerous arrangements of slices along with a shorter lead time. However, they come with other limitations regarding the curvature and layout of the slices. As we will see shortly, the usage of ``sections" is optional and can be ignored if it is unsuitable for the IFU.

\subsection{Parameters used by the DLLs}
\begin{figure}
    \centering
    \includegraphics[width=0.8\linewidth]{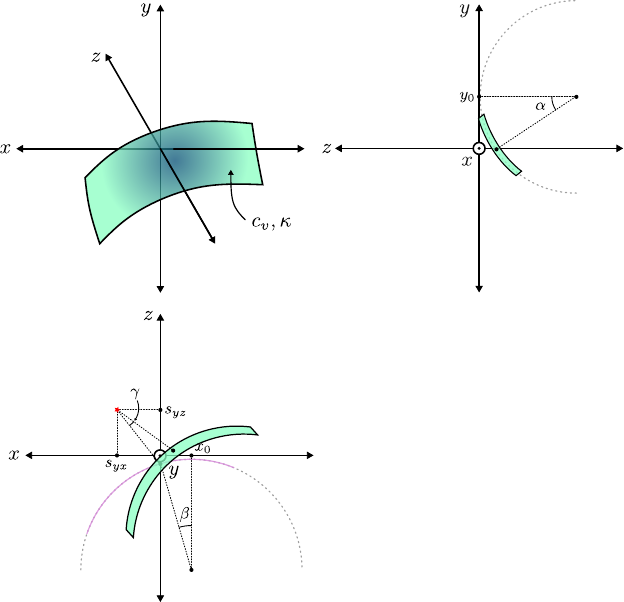}
    \caption{Parameters used to define a single slice. All rotations are about the global coordinate system. $\alpha$ and $\beta$ are depicted as OADs, but they may instead define rotations depending on the surface type. The angle $\gamma$ is separated from $\beta$ as it is used to define rotation of a slice within a section of the image slicer. The slice can be pistoned by an amount $z_p$ after all other transformations have been applied.}
    \label{fig:slice-params-conic}
\end{figure}

As a starting point, consider which parameters are necessary to constrain a single slice. There are several ways to do this, but we will adopt the input parameters in Table~\ref{tab:slice-params}--also illustrated in Figure~\ref{fig:slice-params-conic}--to define the shape and orientation of a slice. The required degrees of freedom are:
\begin{enumerate}
    \item The shape of the untransformed surface $(c_v,\kappa)$.
    \item A tilt of the output ray along the y-axis, either by physically rotating the surface about the x-axis or by applying some off-axis distance (OAD) along the y-axis $(\alpha)$. The y- and z-coordinates of the axis of rotation $(s_{xy},s_{xz})$ if a rotation is used instead of an OAD.
    \item A tilt of the output ray along the x-axis, either by physically rotating the surface about the y-axis and/or by applying some OAD along the x-axis $(\beta,\gamma)$. The x- and z-coordinates of the axis of rotation $(s_{yx},s_{yz})$.
    \item A translation about the z-axis $(z_p)$. (The slicer and pupil mirrors may be translated relative to each other to compensate for field curvature. \cite{henault2004slicing})
\end{enumerate}

\begin{table}[]
\centering
\cprotect\caption{Parameters that describe a single slice.}
\begin{tabular}{|r|p{0.75\linewidth}|}
\hline
Parameter       & Description                                                                                                                                          \\ \hline
$\alpha$        & Off-axis angle along the y-axis OR rotation angle about the x-axis. \\ \hline

$\beta$      & Off-axis angle along the x-axis OR rotation angle about the y-axis.                                                                                                                           \\ \hline
$\gamma$      & Rotation about the y-axis, potentially summed with $\beta$.                                                                                                                                      \\ \hline
$c_v$      & Curvature: $c_v=1/R$ where $R$ is the radius of curvature.                                                                                                                                      \\ \hline
$\kappa$ & Conic constant.                                                                                                           \\ \hline
$z_p$  & Shift along the z-axis (piston).                                                                                                             \\ \hline
$s_{yx}$ & x-coordinate of axis of rotation about y.                                                                                                             \\ \hline
$s_{yz}$ & z-coordinate of axis of rotation about y. \\ \hline
$s_{xy}$ & y-coordinate of axis of rotation about x.              \\ \hline
$s_{xz}$ & z-coordinate of axis of rotation about x.                 \\ \hline


\end{tabular}
\label{tab:slice-params}
\end{table}
The coordinate system is shown in Fig.~\ref{fig:slice-params-conic} with the origin located at the center of the image slicer. The z-axis points in the direction that rays are propagating toward the surface, the y-axis points upward, and the x-axis points left to maintain a right-handed coordinate system. We will describe all rotations as extrinsic transformations, i.e., rotations are performed about the global coordinate axes which do not move with the transformed surface. Analogous parameters for a rotation about the z-axis $(\theta, s_{zx},s_{zy})$ are defined in the source code to allow any possible rigid transformation to be implemented. However, they are irrelevant for our purposes and are thus ignored for the remainder of this paper.

It is more intuitive to think of the parameters $\alpha$, $\beta$, and $\gamma$ as physically rotating the slices. $\alpha$ and $\beta$ define the tilts of the section which determine the x- and y- location of each row of pupil images. In practice, the effective tilt of the section is often defined by an OAD for conicoid surfaces. $\gamma$ then unstacks the slices within a given section and hence always refers to a physical rotation of each slice.

Now that we can define the behavior of individual slices, generating the image slicer entails setting these parameters for each slice plus a few additional parameters to describe the size and layout of the entire surface. As described at the beginning of this paper, the intention of these DLLs is to reproduce the optical behavior of a known image slicer IFU. Thus, a discussion of how to choose these parameters based on system-level requirements will not be explored.

There are two sets of DLLs that determine these parameters using different methods:
\begin{itemize}
    \item \verb|us_slicer_std_ang.dll| and \verb|us_slicer_std_lin.dll|: ``Standard" mode exploits the periodicity of most image slicer designs to minimize the number of parameters that the user interacts with. ``Angular" spaces the parameters evenly in angle while ``linear" sets the parameters so that distances are linearly spaced at a plane downstream of the surface.
    \item \verb|us_slicer_custom.dll|: ``Custom" mode allows the user to manually set the slice parameters in Table~\ref{tab:slice-params} for every individual element.
\end{itemize}
The non-sequential versions of each DLL are \verb|SlicerStdAng.dll|, \verb|SlicerStdLin.dll|, and \verb|SlicerCustom.dll|. Standard mode can be thought of as a convenience wrapper for custom mode since they share a back-end. In standard mode, every slice is assumed to have the same surface type, conic, and curvature; all sections are cut from (potentially different areas of) the same underlying surface shape. See Appendix~\ref{sec:stdparams} for details.

In custom mode, the parameters for individual slices are currently read in from a TXT file. They can be defined to have arbitrary transformations and surface shapes ($c_v$ and $\kappa$), and each row can also be translated by any amount. As a result, there are no sections in custom mode.
With such a large number of degrees of freedom, these DLLs are able to reproduce arbitrary image slicer designs. The user is encouraged to set parameters thoughtfully to avoid generating a surface that is impossible to manufacture. Also note that because parameters are stored in a TXT file, optimization currently cannot be applied to the image slicer in custom mode. In the future, the source code can be modified to pass parameters through ``Data Surfaces" that Zemax can access and modify during optimization. The number of slices is more limited using this method because only 4,800 parameters are available.

In both standard and custom mode, there is a limit on the number of slices because it is impossible to store an infinite number of parameters. The exact number is somewhat arbitrary, but has been set to 5,000 slices, which is just over 500 kB of memory per instance allocated at runtime. Although image slicers rarely exceed a few dozen slices, the limit is significantly higher than this to provide more flexibility while also not using excessive amounts of memory.

The non-sequential DLLs use mostly the same parameters as their sequential counterparts to draw each component as a collection of facets. The facets provide an initial estimate as to where the ray hit the object, upon which an exact solution to the ray trace is provided. The non-sequential DLLs turn the image slicer into a closed 3-dimensional object in a similar fashion to how STL files are created. See Sec.~\ref{sec:nsc} for more information.

\subsection{Brief summary of implementation}
Before discussing results, it would typically be appropriate to include an explanation of how the optical behavior of the surfaces are reproduced by the DLLs. Detailed explanations are provided later in Sec.~\ref{sec:implementation} and \ref{sec:nsc} due to their length. In short, the implementation entails transforming incoming rays out of and back into the global coordinate system according to a transformation matrix specified for each slice, which is defined by the parameters in Table~\ref{tab:slice-params}.

\section{Results and discussion} \label{sec:verification}
\begin{figure}[h]
    \centering
    \includegraphics[width=0.85\linewidth]{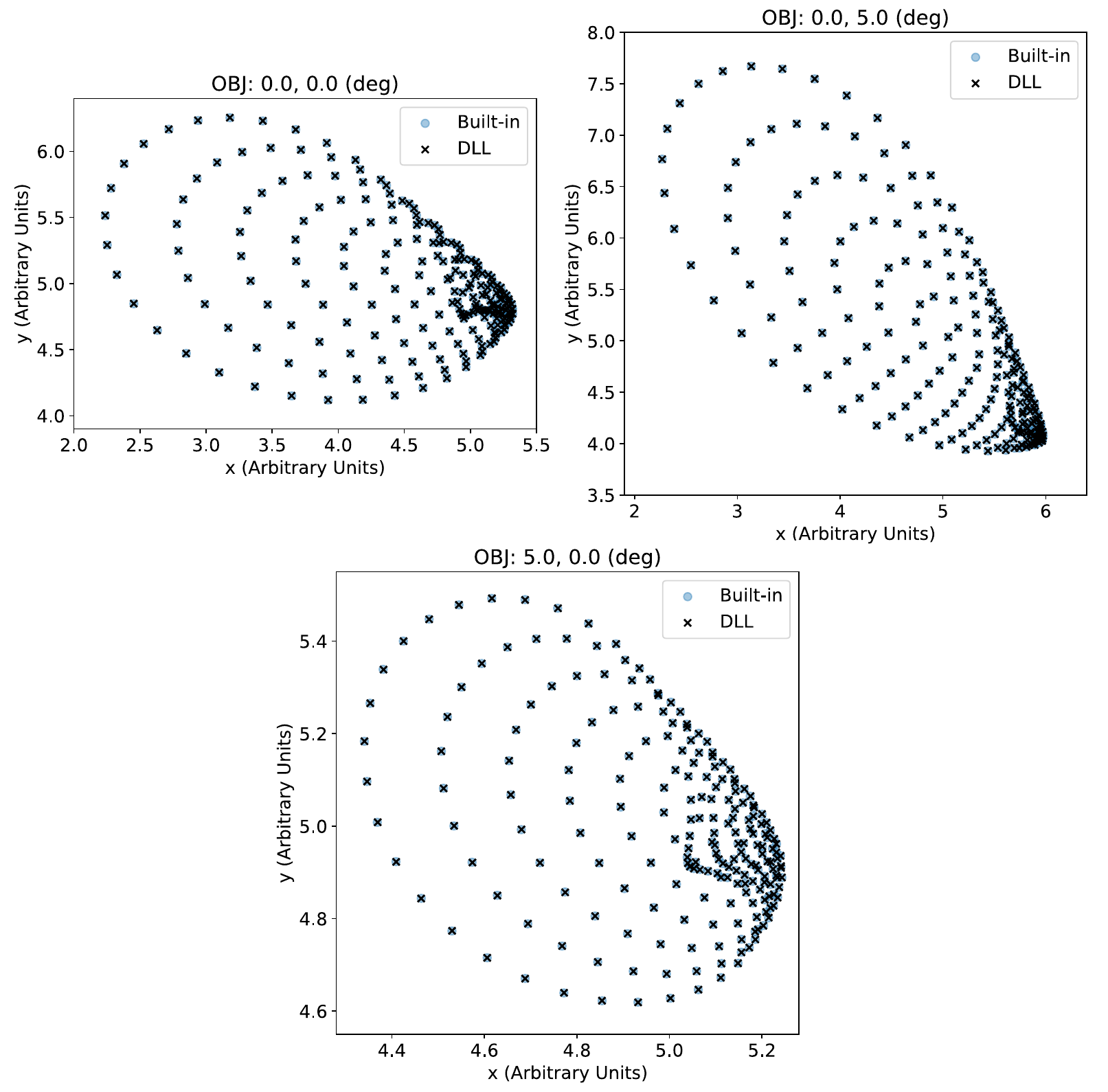}
    \caption{Comparison of spot diagrams generated for a tilted off-axis parabolic mirror. Units on the x- and y-axes are normalized to an arbitrary value for ease of visualization. The surface is set to be the stop and three field angles are set for the object at infinity: \qty{0}{\degree}, \qty{5}{\degree} in y, and \qty{5}{\degree} in x. An \qty{8}{\degree} rotation is performed about the y-axis via coordinate breaks. The OADs $x_0=2.62$ and $y_0=4.37$ are applied to the built-in standard surface using the ``Decenter" option. The ray traces are identical.}
    \label{fig:verify-basic-1}
\end{figure}

The DLLs were validated by checking that they result in identical ray traces compared to using native surfaces in Zemax. Figure~\ref{fig:verify-basic-1} shows a comparison of spot diagrams of a transformed off-axis parabola generated by the DLL versus the built-in ``Standard" surface type, for which the transformations are applied using coordinate breaks. A circular aperture is used on the DLL to match the Standard surface. The spot diagrams show that the ray traces are the same.

\begin{figure}
    \centering
    \includegraphics[width=0.85\linewidth]{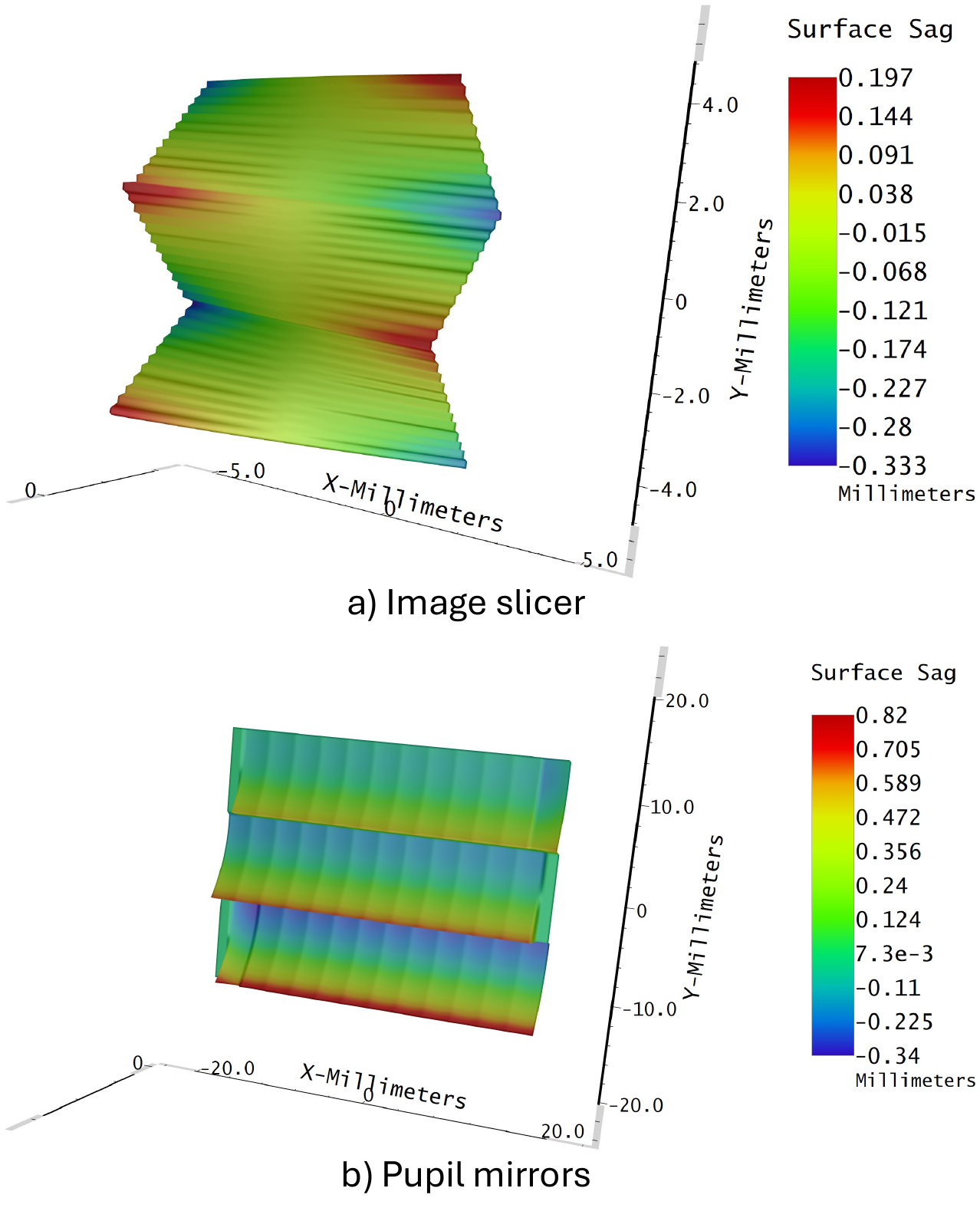}
    \caption{Sag of the SPECTRE's image slicer IFU generated by the DLLs. A rectangle aperture is used to restrict the area of the surface that is drawn. The sag of invalid regions is set to zero for drawing purposes, but are not used for the ray trace.}
    \label{fig:spectre-sag}
\end{figure}

\begin{figure}
    \centering
    \includegraphics[width=0.85\linewidth]{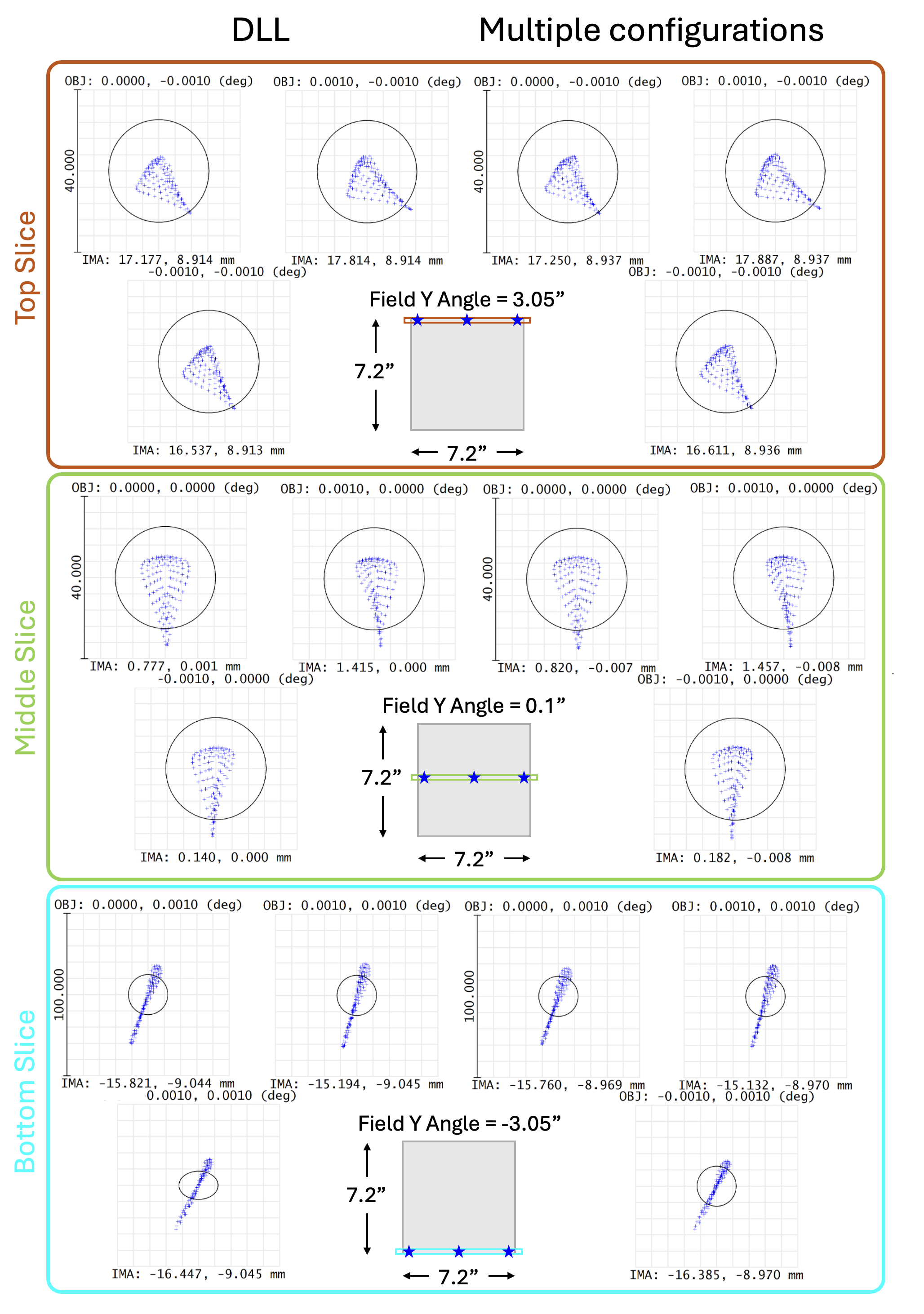}
    \caption{Spot diagrams for selected field angles at the focal plane of SPECTRE's near-infrared spectral channel. The gray square represents the field of view. The rectangles indicates the location of the image plane being sampled along the y-axis and the three blue stars in each rectangle show the sampling across the x-direction. Because the parameters for the DLL were manually adjusted to visually match the existing design, there are negligibly small differences between the two.}
    \label{fig:spectre-spots}
\end{figure}

The DLLs are also capable of reproducing the design of SPECTRE. Figure~\ref{fig:spectre-sag} shows the sag of the image slicer and pupil mirrors modeled by the DLL, and Fig.~\ref{fig:spectre-spots} demonstrates that these surfaces can fully replicate the existing multi-configuration design. There are several major advantages provided by the DLLs. First, SPECTRE contains three separate spectral channels downstream of the reformatted focal plane. Each channel is fed by a dichroic and contains its own prism spectrograph. The DLL allows the channels to be combined into a single file by storing them in separate configurations, so any changes to the foreoptics can be simultaneously applied to all three of them. Next, ray tracing is significantly faster with the DLL compared to using multiple configurations. Although SPECTRE's IFU has 36 slices, the design of the instrument is modified by using files that contain only nine slices because using 36 configurations is prohibitively slow. Representing all 36 slices in the DLL is faster than even the nine slice multi-configuration design, only taking one or two seconds to update the ray trace and basic analysis features.

\begin{figure}[h]
    \centering
    \includegraphics[width=0.85\linewidth]{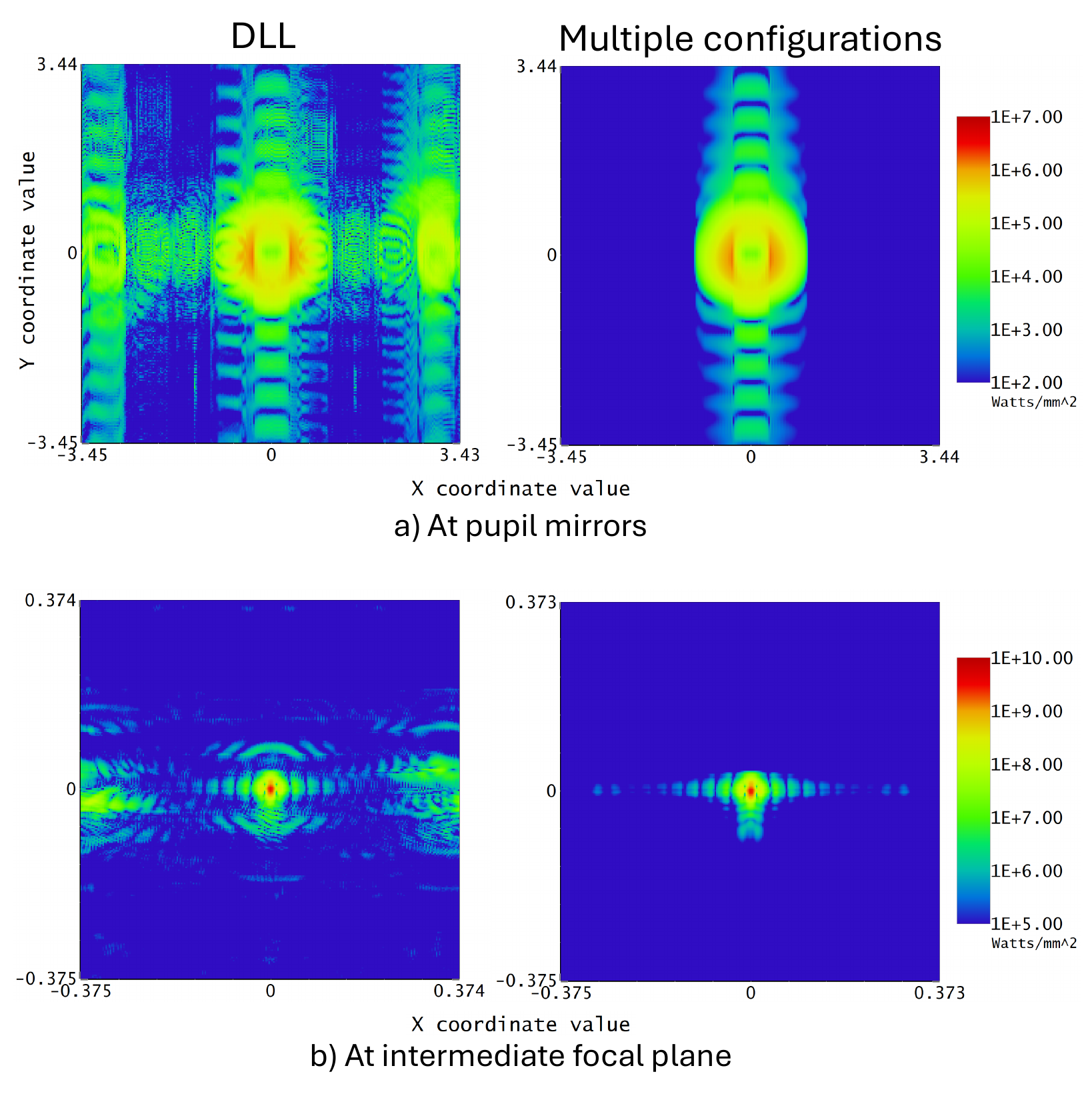}
    \caption{Diffraction in SPECTRE's IFU simulated by physical optics propagation. The intensity is shown on a log-5 scale. Left: The diffraction patterns generated by the DLLs show significant horizontal structure. a) The vertical banding arises from the narrow width of each slice. Light from the Airy rings of the PSF fall on adjacent slices and illuminate their corresponding pupil mirrors. b) Rings of the PSF are re-imaged the intermediate focal plane are shown, along with miscellaneous diffraction from edge effects. The residual diffraction outside of the pseudo-slits is blocked by a baffle. Right: In the original multi-configuration design of SPECTRE, only a single slice can be modeled at a time.}
    \label{fig:pop}
\end{figure}

Finally, it is possible to model effects that cannot traditionally be simulated. Figure~\ref{fig:pop} shows diffraction through SPECTRE's IFU computed using Zemax's Physical Optics Propagation (POP) tool. Using the DLL, one sees diffraction from the outer rings of the Airy point spread function (PSF) that fall on adjacent slices and propagate through to the intermediate focal plane.

\begin{figure}
    \centering
    \includegraphics[width=0.7\linewidth]{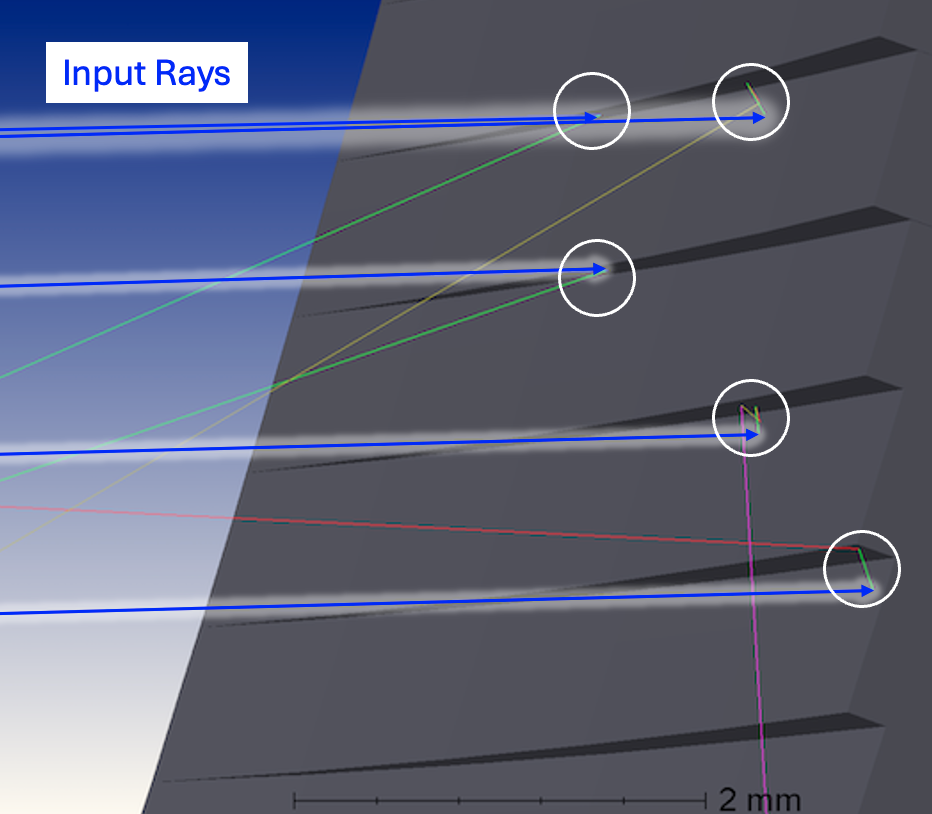}
    \caption{Scattering and reflections off of steps in an image slicer generated by the non-sequential version of the DLLs. The input rays are shown in blue, and each intersection with the surface changes the color of the ray segment. The scattering properties of the slices and walls between slices can be set independently. An unreasonably high value was used for the scattering fraction for demonstration purposes, but this simulation shows that the DLLs can be used to model scattering and grazing incidence reflections.}
    \label{fig:nsc-wallhit}
\end{figure}

The non-sequential DLLs primarily exist to allow the user to quickly convert their sequential design to non-sequential mode using the ``Convert To NSC Group" tool provided by Zemax. Upon using this tool, Zemax will replace the sequential DLL with a ``Grid Sag" surface that does not adequately represent the highly discontinuous surface of the image slicer. The relevant non-sequential DLL can simply be swapped in for this surface and the parameters updated to match the existing design, plus the three extra parameters mentioned at the end of the previous section. Figure~\ref{fig:nsc-wallhit} provides an example of stray light modeled by the DLLs that is caused by light grazing and scattering off of the steps between each slice. It is envisioned that the optical engineer would design their own stray light mitigation using the existing tools in Zemax. For example, a field stop at intermediate focal plane or a pupil mask can be inserted between and around the objects generated by the DLLs to block stray and scattered light. See the stray light analyses for ELT/HARMONI\cite{laurent2018elt} and Gemini/GIRMOS\cite{chabot2022girmos}.

Although image slicers are traditionally merged and re-imported as non-sequential groups, the non-sequential DLLs still provide a few key advantages. The user can easily specify different types of coatings and scatter profiles for surfaces of the slices, steps (walls), and gaps. This can hypothetically be done if the image slicer is described via native non-sequential surfaces or as a CAD file, but is much more tedious. Compared to using a CAD file, the surface can also be represented by a smaller number of elements and still provide a highly accurate ray trace. This reduces the computational burden of drawing the surface and allows more memory to be preserved for tracing and visualizing rays in the non-sequential 3D layout.

The DLLs may fail in extraordinary circumstances which are listed in Appendix~\ref{sec:fail}. Failure modes are related to extremely high curvatures and tilts that cause the surface to become undefined within the bounds of the image slicer.

\subsection{Other applications}
The DLLs can currently be used to describe any rectangular grid of surfaces where each element is either a conic, flat, or cylinder; they can be updated to incorporate new surface shapes without much difficulty as explained in Sec.~\ref{sec:implementation.surftype}. Each element can have a different curvature and may be transformed according to the parameters defined in Table~\ref{tab:slice-params}.

\begin{figure}
    \centering
    \includegraphics[width=0.75\linewidth]{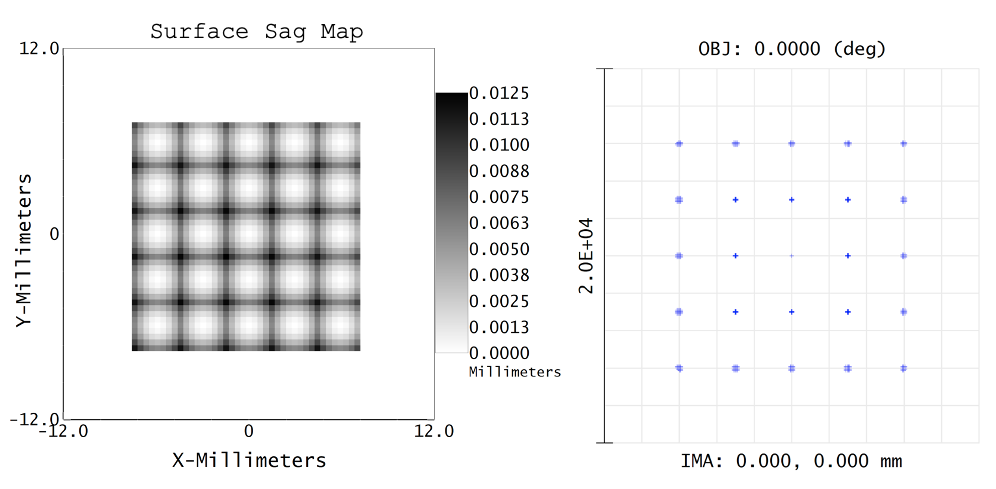}
    \caption{Lenslet array with increasing focal lengths at each ring of lenses generated with custom mode. The sag is shown on the left and the spot diagram at the focus of the central lens is shown on the right. As shown here, the DLL can be used in transmission for other applications.}
    \label{fig:lenslet}
\end{figure}

For example, it is possible to model a microlens array with different focal lengths for each lens (Fig.~\ref{fig:lenslet}). This type of surface appears to have applications in 3D biomedical imaging to capture targets at different depths.\cite{li2022microlens} One may also model a four-sided pyramid wavefront sensor, although it is easier to use a targeted solution such as Ref.~\citenum{antichi2016modeling} which efficiently parametrizes the pyramid. The DLLs may also be used to model a digital micromirror device (DMD, also referred to as a MEMS for micro-electromechanical systems) with an arbitrary number of states.

The DLLs are not suitable for describing segmented telescopes, which are generally composed of a series of hexagonal or circular elements. The processes outlined in Sec.~\ref{sec:implementation} could potentially be generalized to write a DLL that models a segmented telescope aperture. Arbitrary continuous surfaces also cannot be modeled with these DLLs. Is is probably best to do so by reconstructing the surface and using the ZOS-API to either automatically populate the built-in Zernike Sag or Grid Sag surface types.

The sequential DLLs can be modified without much difficulty to accommodate free-form diffraction gratings. This can be done by setting the path length parameter before passing the output ray to Zemax. Conversely, the non-sequential DLLs cannot be used to implement diffractive properties, which must be applied separately to the surface. I have not deeply investigated the necessary functionality for free-form diffraction gratings at the time of writing, but it if the desired properties are not described with built-in features, a diffractive DLL may be implemented in the future to work with the non-sequential DLLs.

\section{Implementation of sequential user-defined surfaces} \label{sec:implementation}
The remainder of this paper is dedicated to explaining the implementation in enough detail that one could fully replicate the DLLs. If one wishes to do so, they are encouraged to look at the source code in the repository. For the average user, the most pertinent section is Sec.~\ref{sec:implementation.sliceparm} which explains how the input parameters of the DLL are converted to the parameters for each slice. We will begin by examining the full implementation for a conicoid surface. This will be followed by a brief explanation of how different surface types are implemented in Sec.~\ref{sec:implementation.surftype}. Appendix~\ref{sec:glossary} provides a description of each variable introduced here.

\subsection{Sag and ray trace for a rotationally symmetric conic} \label{sec:implementation.base}
Within a user-defined surface DLL, Zemax requires a definition for the sag to draw the surface as well as a computation of the ray trace. The latter will be explained momentarily. First, we will briefly review the required equations for an untransformed, rotationally symmetric conic, which is equivalent to the Standard surface in Zemax. The solutions presented here yield the same results as in the \verb|us_stand.c| example provided by Zemax. We will closely follow the procedure in Ref.~\citenum{shannon1997art}. The equations in this subsection are known by many others, but a brief explanation is provided for ease of reference.

The sag of a conic is\cite{shannon1997art}
\begin{equation} \label{eq:conic-sag}
z = \frac{c_v (x^2 + y^2)}{1 + \sqrt{1 - (1+\kappa)c_v^2(x^2+y^2)}}
\end{equation}
where $c_v$ is the curvature, the inverse of the radius of curvature.

To compute the ray trace, it is necessary to 1) parametrize the input ray, 2) calculate where the ray intersects the surface, and 3) compute the refracted or reflected output ray. A ray is defined by the starting coordinates $(x_t,y_t,z_t)$ and direction cosines $(l,m,n)$. The ray is then be transferred to the point of intersection on the surface $(x_s,y_s,z_s)$. That is,
\begin{align}
    x_s &= x_t + t l \label{eq:transfer-1}\\
    y_s &= y_t + t m \label{eq:transfer-2}\\
    z_s &= z_t + tn \label{eq:transfer-3}.
\end{align}
$z_t$ is conventionally defined to be zero, but it will be left as a free parameter for reasons that will become apparent when we transform the surface later. The output ray is defined by the coordinates $(x_s,y_s,z_s)$ and direction cosines $(l',m',n')$. In practice, Zemax provides the input ray parameters (with $z_t$ always set to zero) and expects the DLL to calculate the output ray.

\begin{figure}
    \centering
    \includegraphics[width=0.85\linewidth]{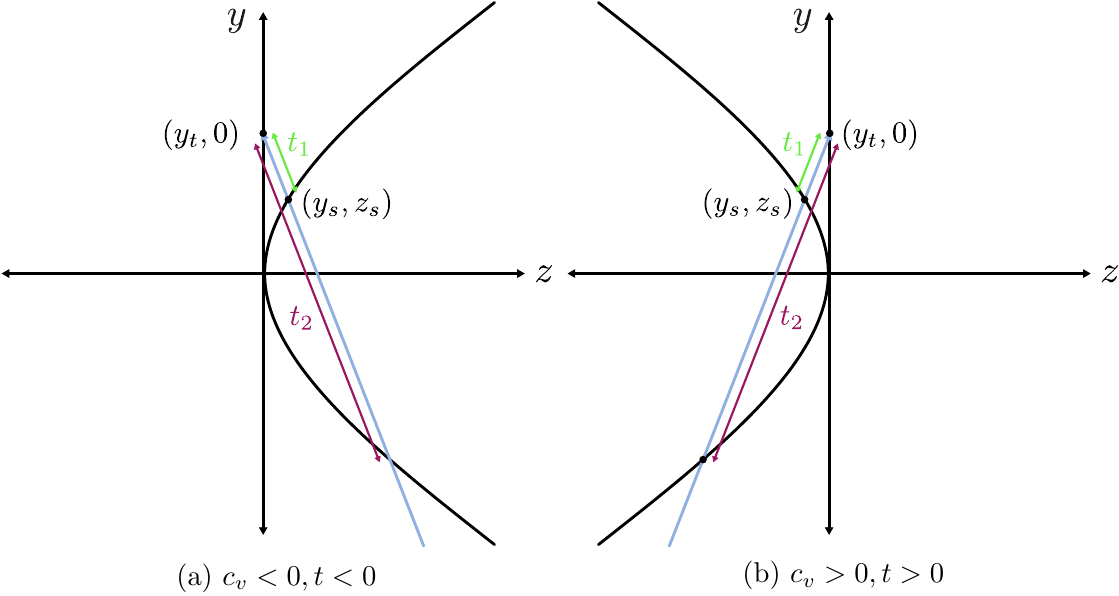}
    \caption{How the sign of $t$ is determined, based on Ref.~\citenum{cheatham1980alternate}. A ray (blue arrow) with two intersection points on a surface, which corresponds to two potential solutions to the transfer distance ($t_1$ in green and $t_2$ in purple). The correct solution for the transfer distance is $t_1$ rather than $t_2$ regardless of whether the surface is concave or convex. Since $|t_1| < |t_2|$, we should pick sign of the root that corresponds to a smaller value of $|t|$. $t$ is generally positive for concave and negative for convex surfaces.}
    \label{fig:t-sign}
\end{figure}

Since we know that $z_s$ is equal to the sag at evaluated at $(x_s,y_s)$, we can use the relation
\begin{equation}
    z_s=\frac{c_v (x_s^2 + y_s^2)}{1 + \sqrt{1 - (1+\kappa)c_v^2(x_s^2+y_s^2)}} \label{eq:transfer-4}
\end{equation}
to solve for $t$ in terms of the input ray's parameters. Substituting \cref{eq:transfer-1,eq:transfer-2,eq:transfer-3} into \cref{eq:transfer-4}, this amounts to solving a quadratic and results in
\begin{equation} \label{eq:rot-conic-1}
    t = \begin{cases} 
        \frac{G}{-F+\sqrt{F^2-DG}} & \text{if } c_v > 0 \\
        \frac{G}{-F-\sqrt{F^2-DG}} & \text{else}
    \end{cases}
\end{equation}
where
\begin{align}
    \begin{split}\label{eq:rot-conic-A}
        D ={}& 1 + \kappa n^2
    \end{split}\\
    \begin{split} \label{eq:rot-conic-B}
        F ={}& x_tl +y_tm + z_tn(1+\kappa) - \frac{n}{c_v}
    \end{split}\\
    \begin{split} \label{eq:rot-conic-C}
        G ={}& x_t^2 +y_t^2 + z_t^2(1+\kappa) - \frac{2z_t}{c_v}.
    \end{split}
\end{align}
The sign of the root may be deduced by considering which side of the surface should be active. Consider a simple case where a ray approaches an unrotated conic surface, shown in Fig.~\ref{fig:t-sign}. It is apparent that we should select the solution that yields a smaller $|t|$.

The surface normal vector is found by taking the gradient of
\begin{equation}\label{eq:sag-surface-for-grad}
    f(x,y,z) = \frac{c_v (x^2 + y^2)}{1 + \sqrt{1 - (1+\kappa)c_v^2(x^2+y^2)}} - z.
\end{equation}
In other words,
\begin{gather}
    \bm{N} = \nabla f \\
    \bm{\hat{N}} = \frac{\bm{N}}{|\bm{N}| }
\end{gather}
where $\bm{N}$ is the surface normal vector and $\bm{\hat{N}}$ is normalized. Evaluating the gradient reveals that the components of the vector are
\begin{align}
    \begin{split}
        N_x &= \frac{\partial f}{\partial x} ={} \frac{c_v x}{\sqrt{1-c_v^2(1+\kappa)(x^2+y^2)}}
    \end{split}\\
    \begin{split}
        N_y &= \frac{\partial f}{\partial y} ={} \frac{c_v y}{\sqrt{1-c_v^2(1+\kappa)(x^2+y^2)}}
    \end{split}\\
    \begin{split}
        N_z &= \frac{\partial f}{\partial z} ={} -1.
    \end{split}
\end{align}

To find the direction of the refracted ray, let
\begin{align}
    \bm{\ell} =& \ l \bm{\hat{x}} + m\bm{\hat{y}}+ n \bm{\hat{z}} \\
    \bm{\ell'} =& \ l' \bm{\hat{x}} + m'\bm{\hat{y}}+ n' \bm{\hat{z}}
\end{align}
so that $\bm{\ell}$ describes the direction of the input ray and $\bm{\ell'}$ the output ray. Also let $\theta_1$ be the angle of incidence, $\eta_1$ the index of refraction in the originating medium, and $\eta_2$ the index in the next medium. The vector form of Snell's law states that\cite{glassner1989introduction}
\begin{equation} \label{eq:snell}
    \bm{\ell'} = \eta_r \bm{\ell} + \Gamma\bm{\hat{N}}
\end{equation}
where
\begin{gather}
    \Gamma =\eta_r \cos\theta_1 - \sqrt{1-\eta_r^2(1-\cos^2\theta_1)} \\
    \cos \theta_1 = -\bm{\ell} \cdot \bm{\hat{N}} \\
    \eta_r = \frac{\eta_1}{\eta_2},    
\end{gather}
which means that the refracted direction cosines are
\begin{align}
    l' &= \eta_r l - \Gamma N_x \\
    m' &= \eta_r m - \Gamma N_y \\
    n' &= \eta_r n - \Gamma N_z.
\end{align}
Reflection is handled by setting $\eta_r=-1$. Zemax provides a \verb|Refract()| function that implements \cref{eq:snell}, computing $\bm{\ell'}$ for a given $\eta_1,\eta_2,\bm{\ell},$ and $\bm{\hat{N}}$.

In the paraxial approximation, the sag is ignored and the ray is defined by its origin $(x_t,y_t)$ and slope angles $(l_s,m_s)$. The direction cosines are converted to slopes using
\begin{align}
    l_s &= \frac{l}{n} \\
    m_s &= \frac{m}{n}
\end{align}
which comes from the small angle approximation. This conversion should always be valid as $n \neq 0$ for paraxial rays. The refracted slope angles are then
\begin{align}
    l_s' &= \frac{1}{\eta_2} \Big( \eta_1 l_s - x_t \Phi_x\Big) \\
    m_s' &= \frac{1}{\eta_2} \Big( \eta_1 l_m - y_t \Phi_y\Big)
\end{align}
where $\Phi_x$ and $\Phi_y$ are the power of the surface along each direction. For a rotationally symmetric conic,
\begin{equation}
    \Phi_x = \Phi_y = (\eta_2-\eta_1) c_v.
\end{equation}
$n'$ is found by converting the refracted slopes to direction cosines with
\begin{align}
    l' = l_s'n' \\
    m' = m_s' n'
\end{align}
and applying the normalization condition $l'^2 +m'^2+n'^2 = 1$ to obtain
\begin{equation}
    n'=\frac{1}{\sqrt{l_s'^2+m_s'^2+1}}.
\end{equation}
Since the surface is approximated by a plane, $z_s=0$ and the normal vector is always
\begin{equation}
    \bm{\hat{N}} = -\bm{\hat{z}}
\end{equation}
regardless of the shape of the surface. The return coordinates will remain unchanged so that $x_s=x_t$ and $y_s=y_t$.

\subsection{Transforming the surface} \label{sec:implementation.transform}
Each slice in the image slicer can be described as a transformed version of the surface sag. The solutions provided in \cref{sec:implementation.base} are valid in the local coordinates of each slice, but need to be transformed into the global coordinate system. This can be done analytically by transforming each of the solutions into global coordinates--shown in \cref{sec:closedform}--but it is easier to transform the input ray into local coordinates, use the existing solutions, and then transform the ray back into global coordinates. The sag can be evaluated as a special case where a ray is traveling parallel to the global z-axis such that $l=m=0$ and $n=1$.

The necessary transformations are parametrized by the variables in Table~\ref{tab:slice-params}. They must be performed in the correct order. For instance, shifting a surface before rotation is very different from applying the shift after the rotation because the former moves the axis of rotation relative to the surface. The order of transformations is determined by the fact that we would like all slices in a given section to be initially polished to the same surface, such that they share all parameters except $\gamma$. The OADs must be applied first. Slices are then rotated about the y-axis. The axis of rotation is offset by applying a translation of $s_{yx}$ and $s_{yz}$, rotating about $y=0$, and translating the surface back by the same amount. Finally, the slice is translated along the z-axis by $z_p$ and it (along with every other slice in the same row) is translated along the x-axis by $u$. For this surface type, $\beta$ denotes an OAD rather than a rotation about the x-axis so $s_{xy}$ and $s_{xz}$ are ignored.

\begin{figure}
    \centering
    \includegraphics[width=0.4\linewidth]{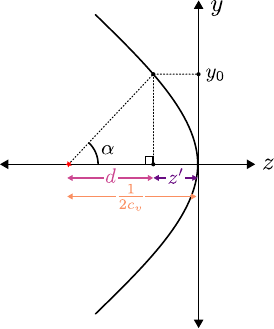}
    \caption{Conversion between the off-axis angle $\alpha$ and the OAD $y_0$. The red star marks the focal point of the conic and $z'$ is the sag at $(0,y_0)$, which is computed using \cref{eq:conic-sag}. $y_0$ can be written in terms of $\alpha$ and $d$ shown in the diagram. The same process can be used to find $x_0$ from $\beta$.}
    \label{fig:off-axis-dist}
\end{figure}

This behavior is described by the matrix $A_{tot}$ where
\begin{equation} \label{eq:transformations-all-conic}
A_{tot} = T R_y T_{OAD}
\end{equation}
\begin{equation} \label{eq:rot-y-gammaonly}
R_{y} =
    \begin{bmatrix}
    1 & 0 & 0 & s_{yx} \\
    0 & 1 & 0 & 0 \\
    0 & 0 & 1 & s_{yz}\\
    0 & 0 & 0 & 1\\
    \end{bmatrix}
    \begin{bmatrix}
    \cos\gamma & 0 & \sin\gamma & 0\\
    0 & 1 & 0 & 0\\
    -\sin\gamma & 0 & \cos\gamma & 0\\
    0 & 0 & 0 & 1 \\
    \end{bmatrix}
    \begin{bmatrix}
    1 & 0 & 0 & -s_{yx} \\
    0 & 1 & 0 & 0 \\
    0 & 0 & 1 & -s_{yz}\\
    0 & 0 & 0 & 1\\
    \end{bmatrix}
\end{equation}
\begin{equation} \label{eq:translation-1}
T =
    \begin{bmatrix}
    1 & 0 & 0 & -u \\
    0 & 1 & 0 & 0 \\
    0 & 0 & 1 & z_p\\
    0 & 0 & 0 & 1\\
    \end{bmatrix}
\end{equation}
\begin{equation} \label{eq:translation-2}
T_{OAD} =
    \begin{bmatrix}
    1 & 0 & 0 & -x_0 \\
    0 & 1 & 0 & -y_0 \\
    0 & 0 & 1 & 0\\
    0 & 0 & 0 & 1\\
    \end{bmatrix}.
\end{equation}
$R_y$ describes a counterclockwise rotation parallel to the y-axis, with the axis of rotation shifted to the coordinates $x=s_{yx}$ and $z=s_{yz}$. The matrices $T$ and $T_{OAD}$ encode different translations, with $T_{OAD}$ being applied before any other transformation and $T$ being applied at the end. $x_0$ and $y_0$ are the OADs determined by $\alpha$ and $\beta$. The OADs along each of the x- and y-axes can be found via Fig.~\ref{fig:off-axis-dist} and are given by
\begin{align}
    y_0 &= \frac{\tan\alpha}{2c_v} \Bigg(\frac{(\kappa-1) + \sqrt{4+\tan^2\alpha(3-\kappa)} }{\tan^2\alpha+(1+\kappa)} \Bigg) \label{eq:off-axis-distance-y} \\
    x_0 &= \frac{\tan\beta}{2c_v}\Bigg(\frac{(\kappa-1) + \sqrt{4+\tan^2\beta(3-\kappa)} }{\tan^2\beta+(1+\kappa)} \Bigg)\label{eq:off-axis-distance-x}.
\end{align}
$\alpha=\ang{90}$ and $\beta=\ang{90}$ are not considered because the optical path would be deflected by \qty{180}{\degree}, which is unphysical. For $\kappa=-1$, these equations reduce to
\begin{align}
    y_0 &= \frac{\sin\alpha}{c_v(1 + \cos\alpha)} \label{eq:oap-y} \\
    x_0 &= \frac{\sin\beta}{c_v(1 + \cos\beta)} \label{eq:oap-x}
\end{align}
which match known definitions for an off-axis parabola.\cite{howard1979imaging} These definitions are insensible if there is no curvature. The equations in Sec.~\ref{sec:implementation.base} may tend to a plane as $c_v \to 0$, but it is more appropriate to re-define the transformation to rotate about both the x- and y-axes instead for planar surfaces; cylinders are handled similarly (Sec.~\ref{sec:implementation.surftype}).

In the linearly spaced version of standard mode, the OADs are provided directly so there is no need for these conversions. The angle of rotation about the y-axis is defined by
\begin{equation}
    \gamma = \arctan(d)
\end{equation}
where $d$ is the spacing between successive increments at a unitary distance along the z-axis from the axis of rotation.

Converting the original surface sag at $(x,y,z)$ the transformed sag at $(x',y',z')$ to involves
\begin{equation} \label{eq:transformed-sag}
    A_{tot} \bm{v} = \bm{v'}
\end{equation}
where
\begin{equation}
    \bm{v} =
         \begin{bmatrix}
            x \\
            y \\
            z \\
            1 \\
        \end{bmatrix}, \;
    \bm{v'} =
         \begin{bmatrix}
            x' \\
            y' \\
            z' \\
            1 \\
        \end{bmatrix}.
\end{equation}
However, transforming the ray into the local coordinates of the slice requires us to invert $A_{tot}$ because in the reference frame of the slice, the transformation is applied backwards. If an input ray is defined by the position $\bm{p_g}$ and direction cosines  $\bm{\ell_g}$ in global coordinates, then it is converted to transformed coordinate system $\bm{p_t}$ and $\bm{\ell_t}$ with
\begin{align}
    A_{tot}^{-1} \bm{p_g} &= \bm{p_t}\\
    A_{tot}^{-1} \bm{\ell_g} &= \bm{\ell_t}
\end{align}
where
\begin{equation}
    \bm{p_g} =
         \begin{bmatrix}
            x_t \\
            y_t \\
            z_t \\
            1 \\
        \end{bmatrix}, \;
    \bm{\ell_g} =
         \begin{bmatrix}
            l \\
            m \\
            n \\
            0 \\
        \end{bmatrix}.
\end{equation}
The fourth term of $\bm{\ell_g}$ is zero to avoid translating the vector, which is necessary to preserve the orientation of the ray encoded by the direction cosines. Although the input ray is defined so that $z_t=0$ in global coordinates, $z_t$ can be non-zero upon transforming the ray into local coordinates, hence why it is retained in \cref{eq:transfer-3}. After computing the output ray in the local coordinate system, it can be transformed back to global coordinates by applying $A_{tot}$.

Note that because $A_{tot}$ describes a rigid transformation, it can be inverted efficiently by decomposing the matrix into a rotation $R$ and translation $\tau$ as such:
\begin{equation}
    A_{tot} = \begin{bmatrix}
    R & \tau \\
    0 & 1 \\
    \end{bmatrix}.
\end{equation}
The inverse is then
\begin{equation}
    A_{tot}^{-1}= \begin{bmatrix}
    R^{-1} & -R^{-1}\tau \\
    0 & 1 \\
    \end{bmatrix}
\end{equation}
where $R^{-1}=R^T$ because rotation matrices are orthogonal (see Chapter III.3 of Ref.~\citenum{heckbert2013graphics}).

\subsection{Generating the image slicer} \label{sec:implementation.sliceparm}
Now that we can compute the behavior of a single slice, we need to define the transformation for every slice to construct the image slicer. In other words, for a given $(x,y)$ position which slice does this correspond do and what are its parameters $(\alpha,\beta,\gamma,z_p,s_{yx},s_{yz},s_{xy},s_{xz})$?

\begin{figure}
    \centering
    \includegraphics[width=0.8\linewidth]{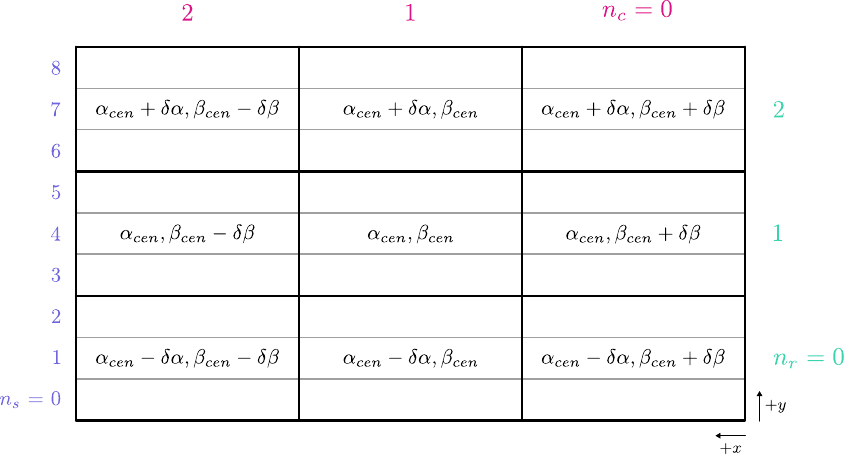}
    \caption{Slicer indices defined in \cref{eq:index-nc,eq:index-ns,eq:index-nr} and corresponding angles $\alpha$ and $\beta$ in standard mode. This diagram shows an image slicer with 9 sections of three slices each viewed face on. $\alpha$ is incremented row-wise while $\beta$ is incremented column-wise.}
    \label{fig:slice-col-idx}
\end{figure}

The number of slices in a column is $n_{slices}=n_{rows} \times n_{each}$. The x- and y-dimensions of the slicer are given by
\begin{align}
    \Delta X&= n_{slices} \delta y + (n_{slices} - 1) g_{y,w}\\
    \Delta Y&= n_{cols} \delta x + (n_{cols} - 1) g_{x,w}.
\end{align}
The column, row, and slice indices can be calculated with floor division as such:
\begin{align}
    n_c &= \lfloor \frac{x - u + \Delta X / 2}{\delta x + g_{x,w}} \rfloor \label{eq:index-nc}\\
    n_s &= \lfloor \frac{y + \Delta Y/ 2}{\delta y + g_{y,w}} \rfloor \label{eq:index-ns}\\
    n_r &= \lfloor \frac{n_s}{n_{each}} \label{eq:index-nr}\rfloor\\
    n_{s,r}&= n_s - n_r \times n_{each} \label{eq:index-nsr}.
\end{align}
$n_c$ is the column index, $n_r$ the row index, $n_s$ the slice index within a column, and $n_{s,r}$ the slice index within the row. Slices within a given $(n_c,n_r)$ correspond to the same section, shown in Fig.~\ref{fig:slice-col-idx}. Differential rotations $\gamma$ are applied to each slice $n_{s,r}$ within a section. Because these expressions yield unreasonable values outside of the defined regions of the image slicer (e.g., $n_r < 0$), they are used to avoid computing the sag when a coordinate $(x,y)$ is out of bounds.

If the user decides to manually define the parameters for each slice using custom mode, the rest of this section is superfluous. The slice parameters for each $(n_c, n_s)$ are instead stored in the array \verb|custom_slice_params|. If each slice is defined in this manner, sections no longer exist because the slices can have arbitrary parameters and are not necessarily polished to match a single surface.

In standard mode, the slice parameters for a given $(n_c,n_s)$ can be determined from the parameters passed to the DLL (Table~\ref{tab:params} in Appendix~\ref{sec:stdparams}). Most of the slice parameters are determined similarly by incrementing along different sections of the image slicer (Fig.~\ref{fig:slice-col-idx}), with the exception of $\gamma$ being incremented along individual slices and $(c_v, \kappa)$ being constant across the entire surface. The parameters $(\alpha, s_{yx}, s_{yz})$ are related to rotations of the output ray about the y-axis and are thus incremented row-wise. Correspondingly, $(\beta, s_{xy}, s_{xz})$ are incremented column-wise. $z_p$ may be incremented both row- and column-wise with the increments $\delta z_{p,row}$ and $\delta z_{p,col}$ being summed together.

We will look at how $\alpha$ is determined as an example. The value of $\alpha$ for the central row depends on whether the number of columns is odd or even. If odd, $\alpha_{cen}$ corresponds exactly to the value of the central row. Otherwise, $\alpha_{cen}$ is the average of the two rows straddling the center $(y=0)$ of the slicer. Thus, for the row index $n_r$ the value of $\alpha$ is
\begin{equation}
    \alpha_{n_r} = \begin{cases} 
        \alpha_{cen} + \delta \alpha ( n_r - \frac{n_{rows} - 1}{2}) & \text{if } n_{rows} \text{ is even}\\
        \alpha_{cen} + \delta \alpha (n_r - \lfloor \frac{n_{rows}}{2}\rfloor) & \text{if } n_{rows} \text{ is odd}.
    \end{cases}
\end{equation}

The angle $\gamma$ is incremented between slices at a spacing of $\delta \gamma$ within a given section. $\gamma_{offset}$ specifies an offset in $\gamma$ between rows. If applied at the focal plane, $\gamma_{offset}$ will cause the resulting rows of pupil images to be horizontally offset relative to each other. This parameter is thus related to the row offset $u$ because a non-zero $\gamma_{offset}$ at the image slicer will necessitate a non-zero $u$ at the pupil mirror array.

\begin{figure}
    \centering
    \includegraphics[width=0.8\linewidth]{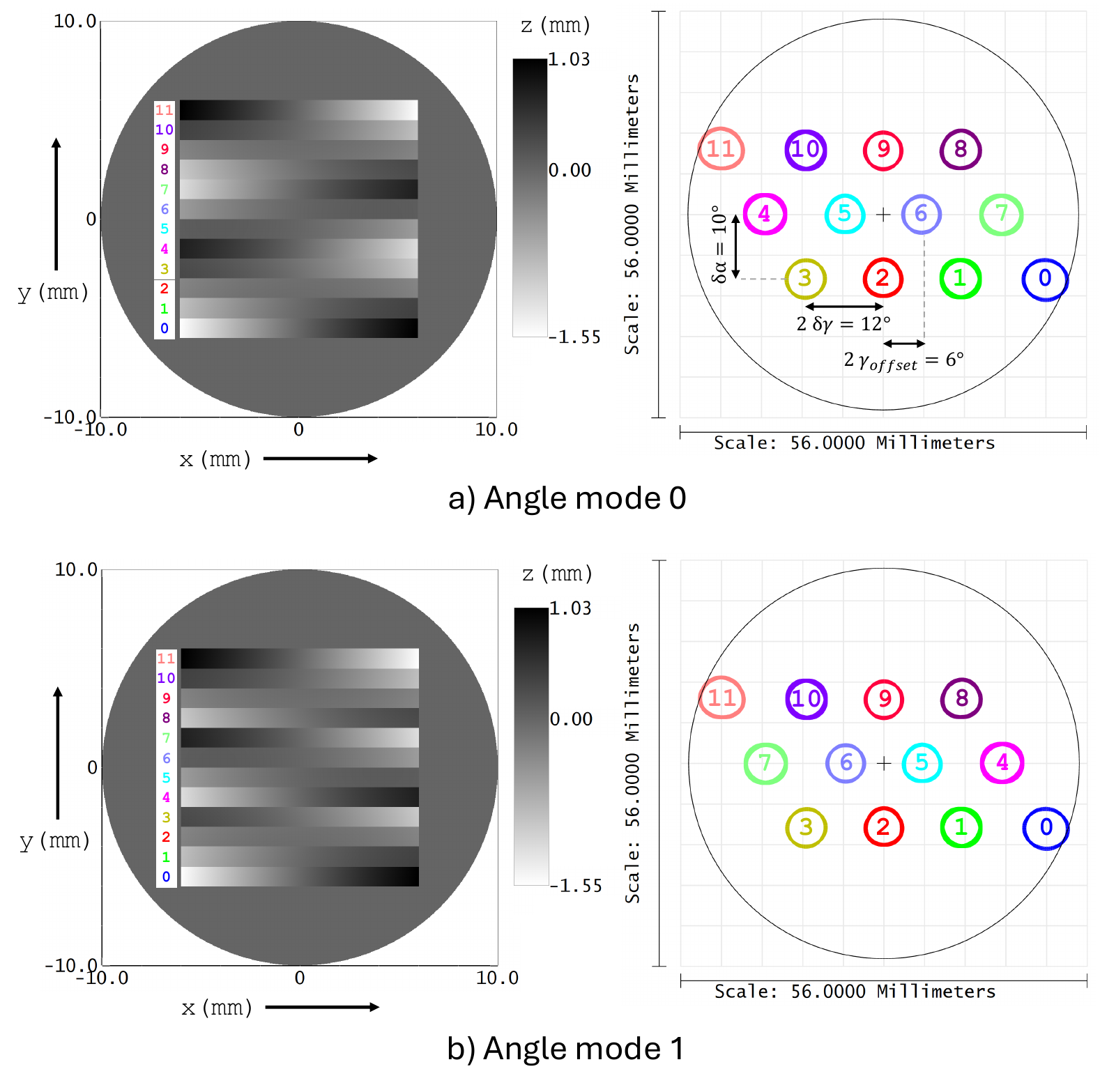}
    \caption{Two of four available angle switching modes. The left shows the surface sag evaluated by Zemax. Invalid regions are displayed as zero. The right shows the footprint diagram at the pupil plane, which shows each of the pupil images created by the 12 slices. This image slicer has three sections of four slices each $(n_{cols}=1,n_{rows}=3,n_{each}=4)$. The rows of pupil images are separated by $\delta\alpha$, which increments the off-axis angle along that defines section. The linear version of the DLL can be used instead to space the images evenly across the pupil plane. a) Modes 0 alternates $\gamma$ in a staircase-like pattern. The difference in angle between each slice within a section is equal to $\delta\gamma$, but the pupil images are separated $2\delta\gamma$ because the surface is used in reflection. Between sections, $\gamma$ is also incremented by $\gamma_{offset}$ which causes the rows of pupil images to be offset from each other. The pupil images snake back and forth because the direction that $\gamma$ is incremented in is flipped between sections. b) Mode 1 does not snake the pupil images because $\gamma$ is incremented identically between sections. This mode is usually not preferred due to large steps that appear between different sections, for example, between slices 3 and 4. }
    \label{fig:angle-modes}
\end{figure}

\begin{figure}
    \centering
    \includegraphics[width=0.8\linewidth]{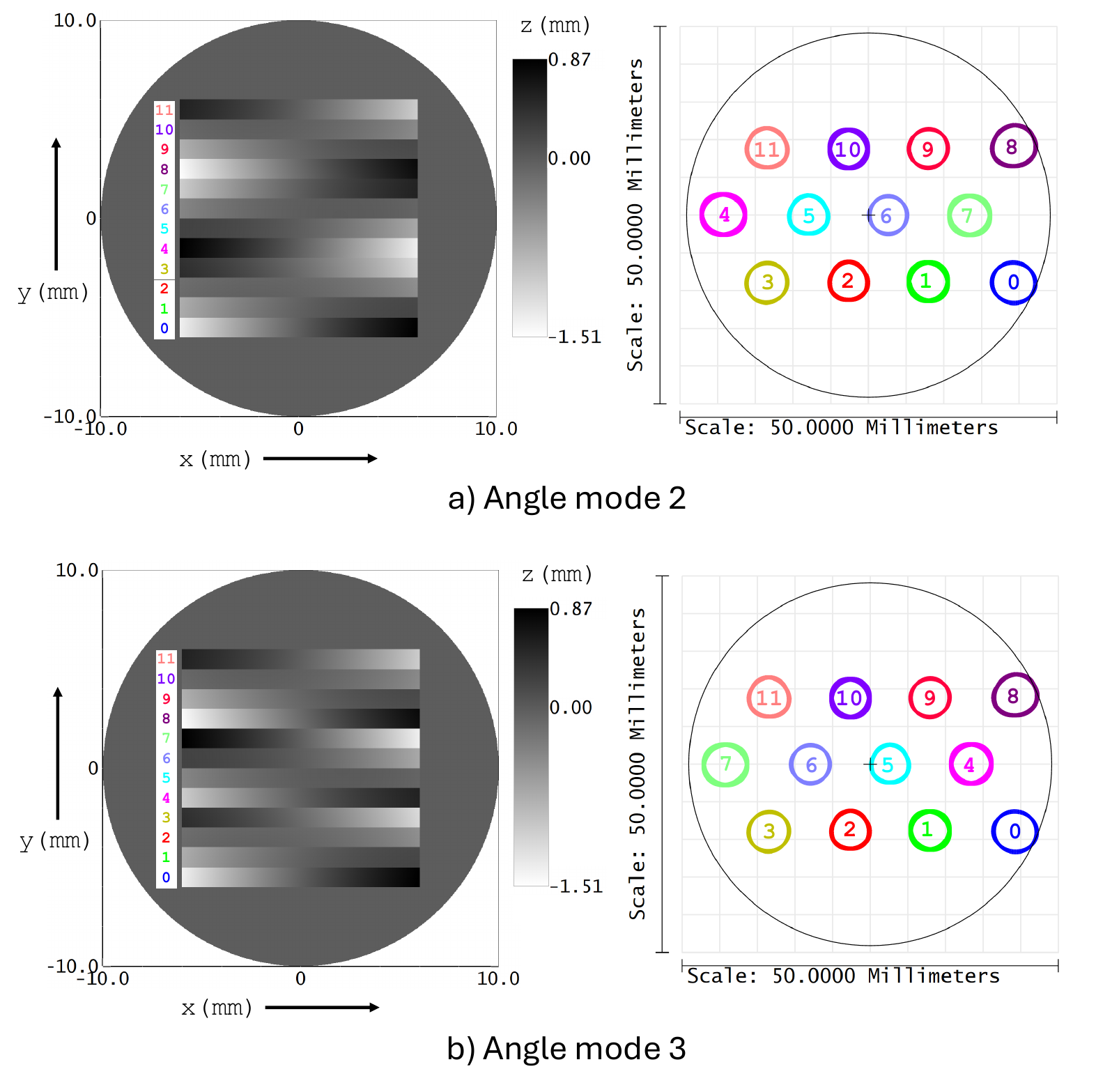}
    \caption{Same as Fig.~\ref{fig:angle-modes} but for modes 2 and 3. a) Mode 2 is similar to mode 0 in that the slices are snaked, but $\gamma_{offset}$ alternated rather than being stacked between rows. This configuration is used in SPECTRE. b) Mode 3 is similar to mode 1 but also alternates $\gamma_{offset}$.}
    \label{fig:angle-modes-2}
\end{figure}

The amount and direction to increment $\gamma$ depends on the specified parameter \verb|angle_mode|. It is easiest to understand how $\gamma$ is incremented visually. Figures~\ref{fig:angle-modes} and \ref{fig:angle-modes-2} demonstrates how different angle switching modes map to the image plane.  Most of the time, the user should select \verb|angle_mode| 0 or 2 to prevent large steps between slices in different sections, which may undesirably increase the amount of stray light. The ``staircase" pattern (\verb|angle_mode = 0|) will compute $\gamma$ with
\begin{align}
\gamma_{bottom}&= \begin{cases} 
        \gamma_{cen} - \delta\gamma \Big( \frac{n_{each}-1}{2} \Big) & \text{if } n_{each} \text{ is even}\\
        \gamma_{cen} - \delta\gamma \lfloor \frac{n_{each}}{2} \rfloor& \text{if } n_{each} \text{ is odd}
    \end{cases} \\
\gamma_{top}&= \begin{cases} 
        \gamma_{cen} + \delta\gamma \Big( \frac{n_{each}-1}{2} \Big) & \text{if } n_{each} \text{ is even}\\
        \gamma_{cen} + \delta\gamma \lfloor \frac{n_{each}}{2} \rfloor& \text{if } n_{each} \text{ is odd}
    \end{cases} \\
\gamma_{extra}&= \begin{cases} 
        \gamma_{offset} ( n_r - \frac{n_{rows} - 1}{2}) & \text{if } n_{rows}  \text{ is even}\\
        \gamma_{offset}  (n_r - \lfloor \frac{n_{rows}}{2}\rfloor) & \text{if } n_{rows} \text{ is odd}
    \end{cases} \\
\gamma&= \begin{cases} 
        \gamma_{bottom} + n_{s,r} \delta\gamma + \gamma_{extra} & \text{if } n_{r} \text{ is even}\\
        \gamma_{top} - n_{s,r} \delta\gamma + \gamma_{extra}& \text{if } n_{r} \text{ is odd}.
    \end{cases}
\end{align}
Other angle switching modes are slight variations of this calculation. One can check the function \verb|GetSliceParamsStandard()| to see how the different modes are implemented.

The parameters of the central slice can be used for the paraxial ray trace. When the number of slices or columns is even, there is ambiguity as to which slice is in the center, so each one can be toggled with via a parameter in the DLL.

\subsection{Ray tracing} \label{sec:implementation.ray}
Zemax provides the input ray and requires us to return information about the output ray. This means that we need to find the transfer distance $t$. $t$ can be used to compute the point of intersection on the surface $(x_s,y_s,z_z)$, which then provides the surface normal and output direction cosines. The problem is that $t$ itself is dependent on the slice parameters at $(x_s,y_s)$, so there is no closed-form relationship between $t$ and the parameters of the input ray. We must find the solution iteratively.

The correct value of $t$ will set $z_s$ equal to the sag of the image slicer. In other words, the computed $t$ should be a root of the following function:
\begin{algorithm}[H]
    \begin{algorithmic}
    \Function{TransferFunction}{$t,x_t,y_t,z_t,l,m,n,\text{p}$}
      \State $x_s = x_t + t l$
      \State $y_s =y_t + t m$
      \State $z_s = t n$
      \State sag = ImageSlicerSag(xs, ys, p)
      \State \Return sag - zs
    \EndFunction
    \end{algorithmic}
\end{algorithm}
\verb|ImageSlicerSag()| computes the sag using the slice parameters which are stored in $p$ in the pseudocode. The piecewise nature of this function and its potential to have multiple solutions makes it poorly suited for most root finding algorithms. A basic approach would be to start at a large value of $t$ and increment inward, checking whether $t$ is a root of \verb|TransferFunction()| at each iteration. This is computationally expensive since many iterations are required to obtain an accurate value of $t$ (within $10^{-12}$). Furthermore, ray intersections with steps (walls) between slices do not register as zeroes of \verb|TransferFunction()| because they are discontinuities in the sag. In sequential mode, correctly tracing the regions between the slices is important for analyzing diffraction and basic stray light analysis.

If we know which slice is struck by the ray, then it is possible to compute an exact solution for the transfer distance. Thus, we need a method to determine which slice to use. Most of the time, the curvature and the angle of incidence of the ray should not be very high. $(x_t,y_t)$ and $(x_s,y_s)$ should then almost always correspond to the same slice except near the boundaries between slices. While we want the ray tracing algorithm to be able to handle all types of rays and image slicers robustly, it should not be much less efficient than if we just assume that $(x_t, y_t)$ corresponds to the same slice as $(x_s, y_s)$. We can apply the following method perform the ray trace:
\begin{enumerate}
    \item Upon initialization of the image slicer, compute the maximum and minimum sag of the entire image slicer, called $z_{max}$ and $z_{min}$. These are called the global extrema. Also compute the maximum and minimum row offsets $u_{min}$ and $u_{max}$. Store these values.
    \item With the given $x_t$, $y_t$, and direction cosines find where the ray intersects the planes $z = z_{max}$ and $z = z_{min}$. Compute the coordinates that the ray intersects each plane at. Use the coordinates to bound the ray.
    \item Determine whether the ray should have intersected the surface based on the computed bounds. If not, the ray missed.
    \item If the ray should have hit, start from $z_{min}$ and iterate through sections.
    \item For each section, determine the number of slices to check and begin iterating through slices.
    \item Compute $t$ for the slice and check if it is a root of the transfer function.
    \item If ray tracing for walls is enabled, check for collisions with walls (and gaps) before iterating to the next slice.
\end{enumerate}

\begin{figure}
    \centering
    \includegraphics[scale=0.95]{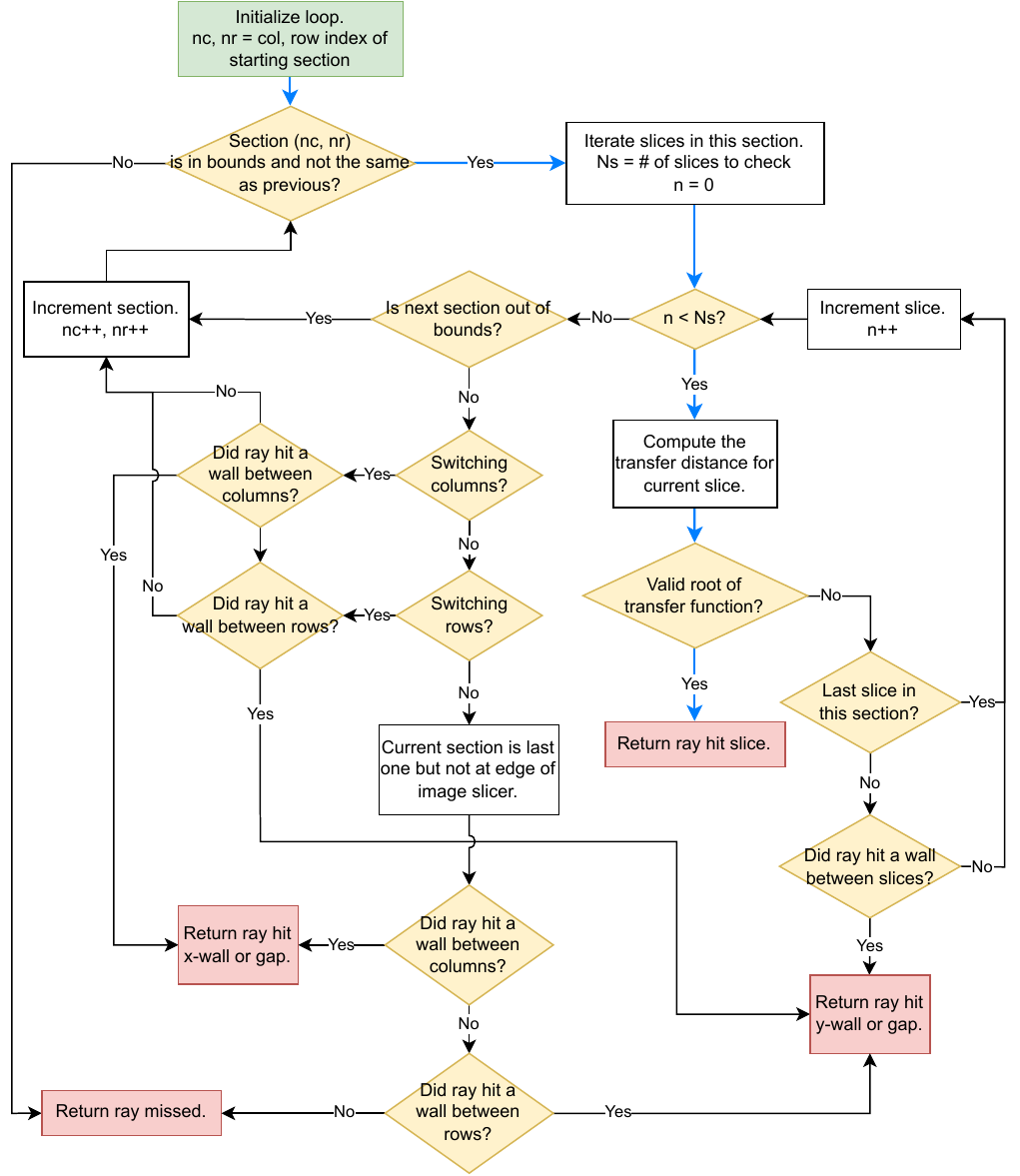}
    \caption{Algorithm for determining which surface is intersected by an incoming ray. This involves propagating the ray from -z to +z and iterating through slices (and adjoining walls) until a valid solution to the ray transfer function is obtained. The vast majority of rays will be confined to a single slice and will thus terminate in one iteration (blue arrows).}
    \label{fig:ray-trace-algorithm}
\end{figure}

\begin{figure}
    \centering
    \includegraphics[width=0.55\linewidth]{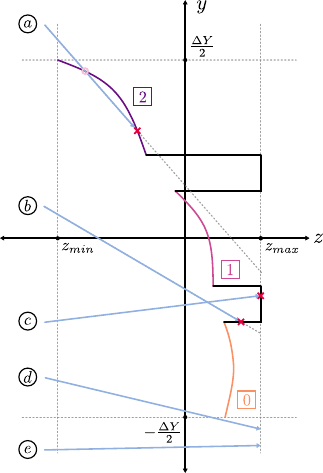}
    \cprotect\caption{Different cases that are handled by the ray tracing algorithm, shown along one axis for ease of visualization. a) The ray hits a slice. Although the line traced by the ray has multiple points of intersection on the image slicer, the ray is considered to have hit the first slice because it is propagated from $z_{min}$ to $z_{max}$. The intersection on the back of the first slice is not considered (see Fig.~\ref{fig:t-sign}). The algorithm will terminate in one iteration without checking the remaining potential solutions. b) The ray hits a wall. The algorithm will check for intersections with slice 1 first, and upon finding no valid solution to the transfer function, will locate the intersection with the wall before checking slice 0. c) The ray hits a gap. Gaps are detected after iterating through every potential slice first and noticing that no slice gives a valid solution of $t$. d) The ray is ``in bounds" at least some of the time between the planes $z_{min}$ and $z_{max}$. In this illustration, $t$ for only the bottom slice is checked before the algorithm determines that the ray missed the slicer. e) The ray--bounded by the planes $z_{min}$ and $z_{max}$--is never in bounds of the image slicer. We immediately know that the ray missed the image slicer without iterating.}
    \label{fig:ray-slicer-cases}
\end{figure}

Figure \ref{fig:ray-trace-algorithm} illustrates this algorithm. The overhead cost is mostly in finding the global extrema, which can be computed upon initialization of the image slicer parameters and recycled for all rays traced through the system. By using the global extrema to find the starting and ending coordinates of the ray, we can quickly determine which slices to check (if any). This avoids iterating through the entire image slicer and correctly handles situations where there are multiple potential solutions for $t$. Starting at the plane $z=z_{min}$, the ray is effectively propagated from -z to +z until the first solution is found. Other solutions cannot be seen by the ray. This has similar behavior to the aforementioned approach of manually incrementing $t$ except that the number of iterations is much smaller, the solutions are exact, and it is possible to check collisions with walls as we increment between slices. The majority of rays will be confined to a single slice within the bounds defined by $z_{min}$ and $z_{max}$ so that the algorithm terminates in one iteration. The algorithm is designed to robustly handle each of the cases shown in Fig.~\ref{fig:ray-slicer-cases}.

To find the global extrema, we will begin by computing a grid of points that coarsely samples the image slicer. From the grid, we can find which slice each extremum is located on. We can then calculate a more accurate solution for the extremum on the slice. For a given slice, the extremum may be at one of the four corners or at the critical point (vertex) of the surface. If the critical point is not located within the bounds of the slice, does not exist, or cannot be computed, then it is excluded and the extremum is found by comparing the sag only at each of the four corners.

The four corners are located at the coordinates
\begin{equation}
    \begin{aligned}
        x_{lo} &= n_c (\delta x+ g_{x,w}) - \frac{\Delta X}{2} + u\\
        x_{hi} &= x_{lo} + \delta x \\
        y_{lo} &= n_s (\delta y + g_{y,w}) - \frac{\Delta Y}{2} \\
        y_{hi} &= y_{lo} + \delta y.
    \end{aligned}
\end{equation}
where the value of $u$ depends on the row index of the current slice. The vertex is located where the partial derivatives of the transformed sag are equal to zero. There will be only one critical point because an unrotated 2D conic has one vertex. For there to be two critical points after rotation, there would have to be multiple values of $z$ for a given $(x,y)$ and the sag would not be a function anymore. We have previously established that the sag is zero for forbidden values of $(x,y)$ to prevent this from occurring.

The value of $y_c$ can be acquired intuitively. The critical point of an axially symmetric conic must be at the origin. Because there is only a single rotation about the y-axis for this surface type, we expect that the critical point will not move in $y$ except for the shift applied via the off-axis distance $y_0$. Therefore,
\begin{equation}
    y_c = -y_0.
\end{equation}
(This may be verified by looking at \cref{eq:dervyC}, the derivative of the analytic solution for the transformed sag.)

The calculation of $x_c$ can be reduced to a one-dimensional problem because we know the value of $y_c$, which allows us to use the secant method of root finding. The derivative is monotonic so it will converge very quickly, often in a single digit number of iterations. Moreover, we should expect the critical point to be near the vertex of the unrotated surface at $x_0$ because the angle of rotation $\gamma$ will usually be low. For the two initial guesses required by the secant method, we can use $x_0 \pm 0.05/c_v$. These values may be undefined if the curvature is very high or angle $\gamma$ is large enough such that the slice is tilted nearly parallel to the z-axis. This should almost never happen, but if it does occur the program will look for the extremum without checking the critical point of the slice. 

Now that we have stored the values of the global extrema, we can compute the bounds of the ray. Using \cref{eq:transfer-1,eq:transfer-2,eq:transfer-3}, the coordinates of the ray at $z=z_{min}$ are
\begin{align}
    t_{min} &= \frac{z_{min}}{n} \\
    x_{min} &= x_t + t_{min} l \\
    y_{min} &= y_t + t_{min} m.
\end{align}
The coordinates at $z=z_{max}$ can be found in the same manner. $n=0$ implies that the ray is traveling perpendicular to the z-axis. This is handled by setting the starting and ending coordinates to the edges of the image slicer
\begin{align}
    x_{min} &= -\frac{\Delta X}{2} + u_{min} \\
    x_{max} &= \frac{\Delta X}{2} + u_{max}
\end{align}
which can be used to solve for $t_{min}, t_{max}, y_{min}$, and $y_{max}$ with \cref{eq:transfer-1,eq:transfer-2,eq:transfer-3}. If $l=0$ as well, then the bounds of the image slicer along the y-axis can be used instead. The minimum and maximum values of $u$ are used to compute the horizontal extent of the image slicer. $u_{min}$ and $u_{max}$ can be pre-computed and stored along with the global extrema.

\begin{figure}
    \centering
    \includegraphics[width=0.8\linewidth]{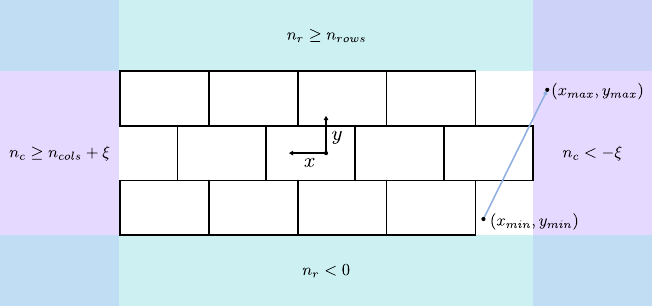}
    \caption{Criteria for determining whether a ray could have hit the image slicer. Let $\xi=\lceil \frac{u_{max} - u_{min}}{\delta x} \rceil$. An image slicer with four columns and three rows (shifted with respect to each other) is shown face-on. A ray is considered out of bounds if, in between $z_{min}$ and $z_{max}$, it is confined to the shaded blue and purple regions, in which case it is guaranteed to have missed the surface.}
    \label{fig:ray-bounds}
\end{figure}

The starting and ending slice indices of the ray can be computed by applying $(x_{min}, y_{min})$ and $(x_{max}, y_{max})$ to \cref{eq:index-nc,eq:index-ns,eq:index-nr}. Refer to the indices of the starting and ending sections as $n_{c,min}, n_{r,min}$ and $n_{c,max},n_{r,max}$, respectively. Before we begin looking for a solution, we can quickly verify whether a ray could possibly have hit the image slicer by looking at these indices. The criteria for whether the ray is always out of bounds are illustrated in Fig.~\ref{fig:ray-bounds}.

If we have determined that the ray is in bounds, then we can begin searching for a solution to the transfer function. This is done with two nested iterations. The algorithm will iterate through sections $(n_c, n_r)$, and for each section, will iterate through each slice until it either finds a solution or reaches the next section. Starting at $(n_{c,min}, n_{r,min})$, let $x_{test} = x_{min}$ and $y_{test} = y_{min}$.

The point of intersection between the ray and beginning of the next column is
\begin{align}
    x_{n_c'} &= \Big( n_c + \frac{1+\zeta_c}{2}\Big)(\delta x + g_{x,w}) - \frac{\Delta X}{2} + u\\
    y_{n_c'} &= y_{test} + \frac{m}{l}(x_{n_c'}-x_{test})
\end{align}
where
\begin{equation}
    \zeta_c = \begin{cases}
    1 & \text{if }  x_{min} \leq x_{max} \\
    -1 & \text{if } x_{min} > x_{max}.
    \end{cases}
\end{equation}
$\zeta_c$ can be thought of as a factor that keeps track of whether the ray is propagating toward +x or -x. It is also used to increment $n_c$. The next row is at
\begin{align}
    y_{n_r'} &= \Big( n_r + \frac{1+\zeta_s}{2}\Big)(\delta y + g_{y,w}) - \frac{\Delta Y}{2} \\
    x_{n_r'} &= x_{test} + \frac{l}{m}(y_{n_r'}-y_{test})
\end{align}
\begin{equation}
    \zeta_s = \begin{cases}
    1 & \text{if }  y_{min} \leq y_{max} \\
    -1 & \text{if } y_{min} > y_{max}.
    \end{cases}
\end{equation}
Whether the next section should be incremented column- or row-wise can be determined by calculating the distance between $(x_{test}, y_{test})$ and the points $(x_{n_c'}, y_{n_c'})$ and $(x_{n_r'}, y_{n_r'})$. If the distance to the former is smaller, then $n_c$ should be incremented and vice versa for $n_r$. The distance to $(x_{max}, y_{max})$ is also calculated; if it is the smallest distance then the current section is the last one to check. Set $x_{next}$ and $y_{next}$ equal to the closest point.

The number of slices to check in the current section is found by applying \cref{eq:index-ns} to $(x_{test},y_{test})$ and $(x_{next},y_{next})$ and taking the difference in the computed $n_s$. Then, setting $x_{test}=x_{next}$ and $y_{test}=y_{next}$ in preparation for the next iteration, we can begin checking whether each slice provides a valid root to the transfer equation.

If the current slice does not yield a valid solution for $t$, we may need to check if the ray collides with a wall before continuing to the next slice. Wall collisions between slices within a column are handled separately from collisions between slices across adjacent columns. The former implies that the wall is parallel to the x-axis while the latter analogously implies that the wall is parallel to the y-axis. Only the former case will be described here, but the implementation is nearly identical in both cases.

\begin{figure}
    \centering
    \includegraphics[width=0.6\linewidth]{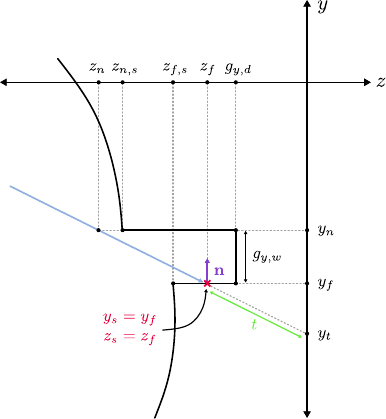}
    \caption{Diagram showing how to determine whether a ray has intersected a wall between slices. The ray will intersect the ``near" wall adjacent to the current slice if the coordinates of the ray at $y=y_n$ lie between $z_{n,s}$ and $g_{y,d}$. Similarly, the ray intersects the wall adjacent to the next slice if the coordinates at $y=y_f$ lie between $z_{f,s}$ and $g_{y,d}$. The ray in this figure intersects the far wall. The surface normal $\mathbf{n}$ at this intersection is shown in purple.}
    \label{fig:ray-wall}
\end{figure}

There may be one or two walls depending on whether the gap size between slices is non-zero. The position of the wall near the current slice is given by
\begin{equation}
    y_n = n_s(\delta y + g_{y,w}) + \Big(\frac{1 + \zeta_s}{2}\Big)\delta y - \frac{\Delta Y}{2}.
\end{equation}
The coordinates of the ray at the plane $y=y_n$ are
\begin{align}
    t_n &= \frac{y_n - y_t}{m}\\
    x_n &= x_t + t_n l\\
    z_n &= t_n n.
\end{align}
Similarly, the coordinates of the ray on the plane defined by the wall across the gap are given by
\begin{align}
    y_f &= y_n + g_{y,w} \zeta_s\\
    t_f &= \frac{y_f - y_t}{m}\\
    x_f &= x_t + t_f l\\
    z_f &= t_f n
\end{align}
The $t_f$ case also handles the situation when the gap size is zero and there is only one wall to check. The criteria for whether a ray has intersected a wall can be determined by examining Fig.~\ref{fig:ray-wall}. Let $z_{n,s}$ be the sag of the nearby slice at $y=y_n$. Correspondingly, $z_{f,s}$ is the sag of the next slice at $y=y_f$. A ray intersects the nearby wall if $z_n$ lies between $z_{n,s}$ and $g_{y,d}$; this is a bit unusual because it can only happen if the gap protrudes out from the slice in the -z direction. A ray intersects the wall adjacent to the next slice if $z_f$ lies between $z_{f,s}$ and $g_{y,d}$. If a ray collision is detected, the transfer distance can be set to either $t_n$ or $t_f$. The surface normal is set to point inward toward the direction of the incoming ray, shown in purple in Fig.~\ref{fig:ray-wall}.

\subsection{Changing surface types} \label{sec:implementation.surftype}
From the previous sections, we can see that only the following four functions are unique to the 2D conicoid surface:
\begin{itemize}
    \item Ray transfer distance
    \item Critical points of the surface (vertices), if applicable 
    \item Gradient of the sag
    \item Transformation matrix given the parameters of the slice
\end{itemize}
Everything else can be recycled if these functions are changed. In practice, different surface types are implemented by choosing an appropriate set of functions depending on the values of $c_v$ and \verb|surface_type|. First, the program checks if the $c_v=0$ and sets the four functions to planar solutions if so. If that is not the case, the program will use the conicoid or cylindrical solutions if \verb|surface_type| is 0 or 1, respectively. New surface types may be implemented by creating the functions listed above and adding a condition to the function that resolves which type of surface to use.

These functions do not necessarily need to be analytic expressions. If an expression for the transfer distance is not readily available, one can either use a root finding method or, if there is a closed form expression for the sag, one can increment $t$ to propagate from $z_{min}$ to $z_{max}$ and check for zeroes of the local transfer function of the surface at each step (refer to the beginning of Sec.~\ref{sec:implementation.ray}). It may be necessary to check the dot product between the surface normal and incoming ray to verify that the ray is hitting the correct side of the surface.

The critical points may also be solved for via iteration as in Sec.~\ref{sec:implementation.ray}, but this may be difficult if the problem cannot be reduced to one dimension because most root finding methods do not work in higher dimensions.\cite{press2007numerical} The easiest workaround would be to set the function for critical points to return NaN values and accept that the calculation of global extrema will be adversely affected. If the surface is guaranteed to have a single vertex, it is possible to instead use the Newton-Raphson method and expect it to converge in two dimensions.

The remainder of this section provides solutions for different surface types implemented by the DLLs. 

\subsubsection{Planes}
Planes have straightforward solutions. The sag is
\begin{equation} \label{eq:flat-sag-base}
    z = 0
\end{equation}
and so the transfer distance is
\begin{equation}
    t = -\frac{z_t}{n}.
\end{equation}
The ray misses if $n=0$ since it is parallel to the surface. The normal vector is
\begin{equation}
    \bm{N} = -\bm{\hat{z}}.
\end{equation}
The direction cosines may be found using the same process as in Sec.~\ref{sec:implementation.base}. For the paraxial ray trace, the surface has no power so
\begin{equation}
    \Phi_x = \Phi_y = 0.
\end{equation}

Define $\alpha$ and $\beta$ to be a tilts about the global y- and x-axes, respectively. The rotation angle about the y-axis is now the sum of $\beta$ and $\gamma$. A series of extrinsic rotations is defined by performing the rotations in the order $R = R_z R_y R_x$. Allowing the axes of rotation to be shifted, the rotation matrices become
\begin{equation} \label{eq:rot-alpha-x}
R_{x} =
    \begin{bmatrix}
    1 & 0 & 0 & 0 \\
    0 & 1 & 0 & s_{xy} \\
    0 & 0 & 1 & s_{xz}\\
    0 & 0 & 0 & 1\\
    \end{bmatrix}
    \begin{bmatrix}
    1 & 0 & 0 & 0\\
    0 & \cos\alpha & -\sin\alpha & 0\\
    0 & \sin\alpha & \cos\alpha & 0\\
    0 & 0 & 0 & 1 \\
    \end{bmatrix}
    \begin{bmatrix}
    1 & 0 & 0 & 0 \\
    0 & 1 & 0 & -s_{xy} \\
    0 & 0 & 1 & -s_{xz}\\
    0 & 0 & 0 & 1\\
    \end{bmatrix}
\end{equation}
\begin{equation} \label{eq:rot-beta-y}
R_{y} =
    \begin{bmatrix}
    1 & 0 & 0 & s_{yx} \\
    0 & 1 & 0 & 0 \\
    0 & 0 & 1 & s_{yz}\\
    0 & 0 & 0 & 1\\
    \end{bmatrix}
    \begin{bmatrix}
    \cos(\beta+\gamma) & 0 & \sin(\beta+\gamma) & 0\\
    0 & 1 & 0 & 0\\
    -\sin(\beta+\gamma) & 0 & \cos(\beta+\gamma) & 0\\
    0 & 0 & 0 & 1 \\
    \end{bmatrix}
    \begin{bmatrix}
    1 & 0 & 0 & -s_{yx} \\
    0 & 1 & 0 & 0 \\
    0 & 0 & 1 & -s_{yz}\\
    0 & 0 & 0 & 1\\
    \end{bmatrix}
\end{equation}
and including the ability to piston sections and offset rows with respect to each other,
\begin{equation} \label{eq:flat-transformation}
A_{tot} = TR_yR_x.
\end{equation}
$T$ is given by \cref{eq:translation-1}. Note that because $\alpha$ and $\beta$ now define rotations of the surface, the output angle of the optical axis is doubled compared using the OAD definitions in \cref{eq:off-axis-distance-x,eq:off-axis-distance-y}.

\subsubsection{Cylindrical surfaces}
For cylindrical surfaces that follow the shape of a conic along the x-axis but have no curvature along the y-axis, the sag is
\begin{equation}\label{eq:sag-cylinder}
    z = \frac{c_v x^2}{1 + \sqrt{1 - (1+\kappa)c_v^2x^2}}.
\end{equation}
The transfer distance and surface normals are comparable to the solutions for the rotationally symmetric conic:

\begin{equation}
    t = \begin{cases} 
        \frac{G}{-F+\sqrt{F^2-DG}} & \text{if } c_v > 0 \\
        \frac{G}{-F-\sqrt{F^2-DG}} & \text{else}
    \end{cases}
\end{equation}
\begin{align}
    \begin{split}
        D ={}& 1 + \kappa n^2
    \end{split}\\
    \begin{split} 
        F ={}& x_tl + z_tn(1+\kappa) - \frac{n}{c_v}
    \end{split}\\
    \begin{split}
        G ={}& x_t^2 + z_t^2(1+\kappa) - \frac{2z_t}{c_v}
    \end{split}
\end{align}
\begin{align}
    \begin{split}
        N_x &= \frac{c_v x}{\sqrt{1-c_v^2(1+\kappa)x^2}}
    \end{split}\\
    \begin{split}
        N_y &= 0
    \end{split}\\
    \begin{split}
        N_z &= -1.
    \end{split}
\end{align}
The surface is only curved about the x-direction, so for the paraxial ray trace the power is
\begin{align}
    \Phi_x &= (\eta_2-\eta_1) c_v \\
    \Phi_y &= 0.
\end{align}

The angle $\alpha$ is used to define a rotation about the global x-axis, but the angle $\beta$ will be used to define an OAD along the x-axis of the conic. The angle $\gamma$ still defines a rotation about the global y-axis. The rotation matrices $R_x$ and $R_y$ are given by \cref{eq:rot-alpha-x} and \cref{eq:rot-y-gammaonly}. The OAD $x_0$ (\cref{eq:off-axis-distance-x}) is applied prior to any rotations, so
\begin{equation} \label{eq:cylinder-transformation}
    A_{tot}=TR_yR_x T_{OAD}
\end{equation}
where
\begin{equation}
    T_{OAD}=\begin{bmatrix}
    1 & 0 & 0 & -x_0 \\
    0 & 1 & 0 & 0 \\
    0 & 0 & 1 & 0\\
    0 & 0 & 0 & 1\\
    \end{bmatrix}.
\end{equation}
We can use the definition of $T$ in \cref{eq:translation-1} here as well.


\section{Implementation of non-sequential user-defined objects} \label{sec:nsc}
Non-sequential mode is often used to perform more advanced stray light analyses than can be done in sequential mode. Because sequential DLLs cannot be used in non-sequential mode, it is advantageous to create dedicated non-sequential DLLs that can generate identical surfaces using the same parameters.

To create the non-sequential object, Zemax requires that the surface is approximated by a series of triangles, comparably to how objects are represented by STL files. Zemax uses these triangles to draw the object and obtain an initial guess as to where the ray intercepts the surface. Then, we must provide an exact solution to the ray transfer distance and surface normals. Generating the triangles is laborious but not difficult given that we have already written a function for the sag. Additionally, Zemax reports the $(x,y)$ coordinate of intersection for the ray and also allows us to flag which type of surface it hit. This simplifies the ray tracing process substantially as we can can immediately solve for the transfer distance without the algorithm described in Sec.~\ref{sec:implementation.ray}.

The same parameters are used as for the sequential DLLs, plus three more:
\begin{itemize}
    \item $N_x$: Number of facets along the x-direction to represent a slice.
    \item $N_y$: Number of facets along the y-direction.
    \item $Z_{diff}$: Distance from the global maximum of the image slicer to the rectangular shell at the back of object. (The back is located at the largest z-value, hence the global maximum rather than minimum.)
\end{itemize}

\subsection{Triangle generation}
\begin{figure}[h]
    \centering
    \includegraphics[width=0.7\linewidth]{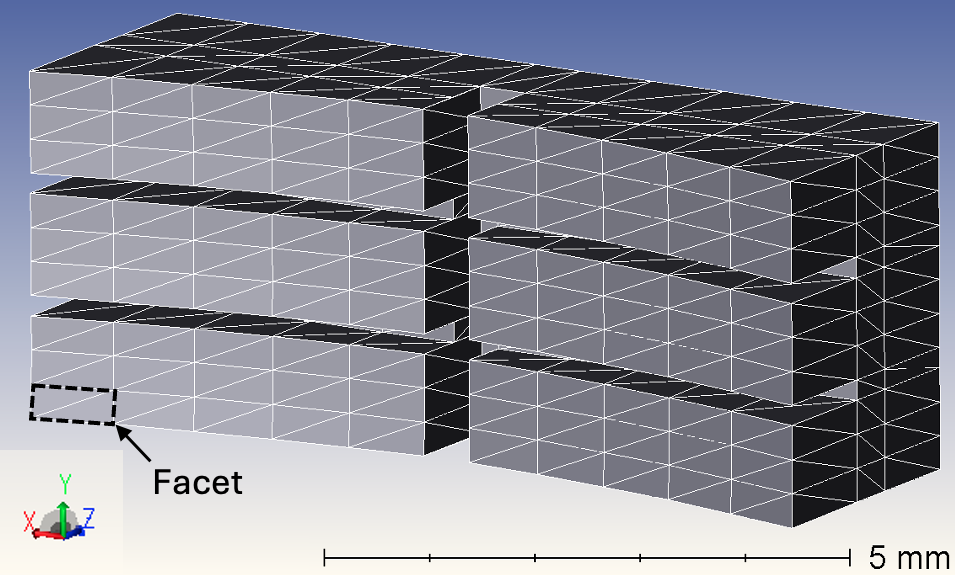}
    \caption{Surface generated by one of the non-sequential DLLs. The surface is composed of a series of rectangular facets (a single facet is outlined in black). The number of facets may be defined by the user. The curvature and dimensions of the slices are exaggerated to aid in visualization. The object shown in this figure has three rows, two columns, and gaps between the slices along both x and y. The gap widths are set to non-zero values for illustration purposes.}
    \label{fig:nsc-facets}
\end{figure}

\begin{figure}[h]
    \centering
    \includegraphics[width=0.8\linewidth]{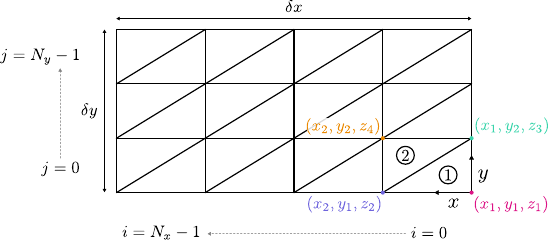}
    \caption{Slice approximated by twelve facets $(N_x=4,N_y=3)$. Each facet is composed of two triangles. Prior to generating the triangles, a grid of $(x,y,z)$ coordinates for every slice is calculated and stored. The triangles are then generated by iterating through every facet on the grid and selecting the coordinates at the corners of each facet. Then, those coordinates are sent to a function that will create two triangles for that facet. For example, the facet defined by the four points $p_n = (x_n,y_n,z_n)$ where $n=1\text{ to }4$ will result in triangles with vertices $p_1,p_2,p_3$ and $p_2,p_3,p_4$.}
    \label{fig:tri-slice}
\end{figure}

\begin{figure}[h]
    \centering
    \includegraphics[width=0.4\linewidth]{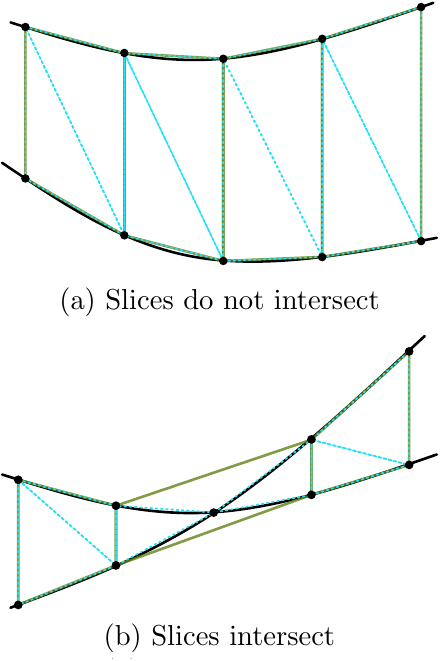}
    \caption{Triangles for walls between slices. The facets (dark green boxes) approximate the step between the two slices (black curved lines). If the slices do not cross as in a), the facets can be generated as normal. However, if the slices cross within the facet as in b), the point of intersection is evaluated and used as a vertex for the triangles. If the point of intersection is on the edge of a facet, only one triangle is generated.}
    \label{fig:tricross}
\end{figure}

The DLLs allow the user to specify the number of rectangular facets (Fig.~\ref{fig:nsc-facets}) used to approximate each slice along the x- and y-directions. Each rectangle is composed of two triangles. The appropriate number of facets depends on the optical system, but if the curvature is low and the user does not expect rays to come in from a very high angle of incidence, only one or two facets are needed along each direction. Since we can provide an exact solution to the ray trace upon being provided which triangle was struck by the ray, the sampling only needs to be good enough that a ray will not intersect the wrong triangle. It is helpful to identify four different groups of surfaces on the object:
\begin{itemize}
    \item Surfaces of slices
    \item Walls (steps) between slices
    \item Gaps between slices
    \item Outer shell of the surface
\end{itemize}
Different coating and scattering properties can be applied to each of these four groups. Generating the triangles involves evaluating the sag along the grid defined by $N_x$ and $N_y$ (Fig.~\ref{fig:tri-slice}) to create the surface of each slice. Then, facets are drawn between the slices and a shell is added to enclose the object. (The grid functions even when the rows of slices are horizontally offset from each other when $u>0$.) Each facet is checked for crossover points (Fig.~\ref{fig:tricross}) to draw overlap regions between slices correctly. If the reader wishes to know how exactly all of the triangles generated, they may refer to the source code in \verb|triangles.c|.

\section{Conclusion}
I have developed a set of DLLs that can model image slicer integral field units in both the sequential and non-sequential ray tracing modes of Zemax. The DLLs are able to accurately replicate analysis results from natively transformed surfaces in Zemax and have replicated the design of SPECTRE, a facility spectrograph for NASA IRTF. The DLLs provide a variety of advantages over the traditional method of designing image slicers, including a drastic increase in ray tracing speed, the ability to fully represent an instrument with many slices and/or multiple spectral channels in a single file, and the ability to model diffraction from the entire IFU which cannot be simulated in Zemax otherwise.

The DLLs can also be used to generate a nearly arbitrary grid of surfaces and are written to be easily extensible to new types of surfaces that have not been implemented yet. The methods introduced in this paper may be of interest in any scenario where it is desirable to efficiently trace rays through a discontinuous mesh of surfaces.

Recalling the problems described in Sec.~\ref{sec:intro.challenges}, the primary motivation of this work was to improve Zemax's ability to simulate image slicer IFUs. In the future, it is necessary to develop additional design tools that will assist in generating a prescription for the IFU starting from system-level requirements. In principle, the requirements mentioned in Sec.~\ref{sec:intro.challenges} should deterministically affect the desired layout of the image slicer IFU. Thus, the process of generating parameters for the IFU should be automated as much as possible, after which they can be passed to the DLLs described in this paper. Furthermore, it should be possible to generate human-readable CAD files--for example, with the SolidWorks API--that can be used to specify image slicers for vendors.

Regarding short-term improvements to the DLLs: they should be modified to take parameters via a data surface to allow optimization of the image slicer. Additionally, the parameters used in standard mode may not ideal when there are multiple rows of pseudo-slits (as in SPECTRE) because the spacing of angles between slices are not necessarily the same in each section. Optimization in Zemax is also not designed to accommodate reformatting the image plane, although it is still possible to set the built-in operands in a manner to correctly optimize the system. It is necessary to further explore the most effective manner in which to use these DLLs.
\appendix

\section{Potential failure modes} \label{sec:fail}
There are a few edge cases that may cause the DLLs to fail, which are mostly related to extremely high curvatures or inappropriate usage of the surface. These should not be an issue in the vast majority of cases. There are no strict limits on the radius of curvature, but generally speaking it should not approach the length and width of each slice to avoid extremely sharp edges. If it is so large that the slices are basically flat $(c_v < 10^{-13})$, the radius should be set to "Infinity" in Zemax to avoid using the conicoid solutions. Other invalid parameters (e.g., a negative slice width) are handled automatically by the DLLs.

Known failure modes are listed here for completeness.
\begin{itemize}
    \item The curvature is too high to obtain a good estimate of the global extrema: The global maximum and minimum are necessary to compute the ray trace as described in Sec.~\ref{sec:implementation.ray}. These values are computed by first performing a coarse sampling of the image slicer to determine which slice the extremum is located on, and then by computing a more accurate (within $10^{-13}$) value for the extremum on that slice. The former step may fail if the radius of curvature near the width of a slice. This problem can be ameliorated by increasing the sampling frequency for finding the global extrema or by using an ad hoc estimate for the global extrema. Either would require modifying the source code. \
    \item For a 2D conic, $\alpha$ and $\beta$ are non-zero when the curvature is extremely small $(c_v<10^{-13})$ but non-zero: this will not cause the program to crash, but the result is likely to be invalid because the OAD goes to infinity $c_v$ goes to zero. Planar surfaces are implemented separately so that $\alpha$ and $\beta$ describe rotations about the y- and x-axes, respectively. Set the radius of curvature to $\infty$ in Zemax to avoid this problem.
    \item Rays are traced to the back side of the mirror array: The user has somehow flipped the entire surface front-to-back such that it is being used backward. If enabled, the method of detecting wall collisions for the sequential ray trace (Sec.~\ref{sec:implementation.ray}) depends on the surface being used front-to-back. Intersections with walls may not be detected appropriately if the rays approach the surface from the wrong side. The surface should not be used in this manner. If a convex array of surfaces is desired then the user can set the curvature to a positive number instead.
    \item For the non-sequential DLLs, if the number of facets used to draw the surface does not sample the curvature of the slices well enough: The non-sequential DLL requires that the image slicer is represented as a series of triangles. Zemax uses these triangles to determine where the ray was intersected, and then asks the DLL to provide the exact solution for the ray trace. The number of triangles (facets) used to draw the slicer can be specified by the user. If the sampling is too coarse, then a ray can potentially hit the wrong facet. This problem is fundamental to how Zemax's user-defined object DLLs are implemented. To avoid this issue, the user should set the number of facets to adequately sample the curvature the slices. This is not a problem for flat slices because they are represented exactly by triangles.
\end{itemize}
I have attempted to carefully avoid divisions by zero, segmentation faults, or other errors that may cause Zemax to crash. However, Zemax does not perform safety checks on user-defined DLLs to improve computational speed. Additional work is needed to ensure that the DLLs are very robust to prevent the user from losing data if Zemax crashes.

\section{Parameters in standard mode} \label{sec:stdparams}
Table~\ref{tab:params} describes the parameters used to manipulate the image slicer IFU in standard mode. The number of sections is determined by $n_{rows}$ and $n_{cols}$, with the number of slices per section being $n_{each}$. The overall dimensions of the image slicer are determined by these parameters along with $\delta x$ and $\delta y$, which are the length and width of a single slice. Within a given section, all slices are assumed to have the same ``base" surface shape and axes of rotation. Many of the parameters in Table~\ref{tab:params} are used to determine how to increment variables between different sections. Precise definitions are provided in Sec.~\ref{sec:implementation.sliceparm}.

Some parameters may go unused depending on the surface type. For example, the parameters $s_{xy}$ and $s_{xz}$ are not used for the rotationally symmetric conic because there is only a single rotation about the y-axis. The user may also specify the width and depth of moats (called "gaps" in this paper) between each slice using the parameters $g_{x,w},g_{x,d},g_{y,w},$ and $g_{y,d}$. These are generally not used for image slicers but are included for compatibility with more kinds of surfaces. If the gap widths are set to zero then the depths will not be used.

\begin{table}[]
\centering
\cprotect\caption{Parameters that describe an image slicer in standard mode. Variables in parentheses are used for the linear version of standard mode which, for example, would space the subpupil images produced by the image slicer evenly across intermediate pupil plane (as opposed spacing them evenly in angle). If these parameters are insufficient, the user may use custom mode to define the parameters for each individual slice.}
\begin{tabular}{|r|p{0.75\linewidth}|}
\hline
Parameter       & Description                                                                                                                                          \\ \hline
\verb|surface_type|        & If 0, surface is a rotationally symmetric conic. If 1, a cylindrical conic. \\ \hline

$n_{each}$      & Number of slices per section.                                                                                                                          \\ \hline
$n_{rows}$      & Number of rows of sections.                                                                                                                                      \\ \hline
$n_{cols}$      & Number of columns of sections.                                                                                                                                     \\ \hline

\verb|angle_mode|            & Pattern generated by $\gamma$ (see Sec.~\ref{sec:implementation.sliceparm}).                                                                                                     \\ \hline
$\delta \alpha$ (or $\delta y_0$) & Change in off-axis angle between rows in degrees.                                                                                                           \\ \hline
$\delta \beta$  (or $\delta x_0$)& Change in off-axis angle between columns in degrees.                                                                                                             \\ \hline
$\delta \gamma$  (or $\delta d$)& Change in rotation angle about the y-axis between slices within a section in degrees.                                                                                                             \\ \hline
$\gamma_{offset}$ (or $d_{offset}$)& Offset in $\gamma$ between rows; misalignment of rows of subpupil images with respect to each other. \\ \hline
$\alpha_{cen}$  (or $y_{0,cen}$)& If $n_{each}$ is odd: the angle of the central row in degrees. Else: mean of the angle between the central 2 rows. \\ \hline
$\beta_{cen}$   (or $x_{0,cen}$)& Angle of the central column in degrees; analogous to $\alpha_{cen}$ but for $\beta$.                                                                 \\ \hline
$\gamma_{cen}$   (or $d_{cen}$)& Angle of the central slice in degrees; analogous to $\alpha_{cen}$ but for $\gamma$.                                                                 \\ \hline
$z_{ps,cen}$  & Translation along z of the central slice in a section. \\ \hline
$z_{p,cen}$  & Translation along z of the central section. \\ \hline
$s_{yx,cen}$ & x-coordinate of axis of rotation about y for the central slice. \\ \hline
$s_{yz,cen}$ & z-coordinate of axis of rotation about y the central slice. \\ \hline
$s_{xy,cen}$ & y-coordinate of axis of rotation about x for the central slice, if applicable. \\ \hline
$s_{xz,cen}$ & z-coordinate of axis of rotation about x for the central slice, if applicable. \\ \hline
$u_{cen}$    & Row offset along the x-axis for the central slice. \\ \hline
$\delta z_{ps}$ & Change in $z_p$ slices within a section.              \\ \hline
$\delta z_{p,col}$ & Change in $z_p$ columns.              \\ \hline
$\delta z_{p,row}$ & Change in $z_p$ between rows.                 \\ \hline
$\delta s_{yx}$  & Change in $s_{yx}$ between rows. \\ \hline
$\delta s_{yz}$  & Change in $s_{yz}$ between rows. \\ \hline
$\delta s_{xy}$  & Change in $s_{xy}$ between columns, if applicable. \\ \hline
$\delta s_{xz}$  & Change in $s_{xz}$ between columns, if applicable. \\ \hline
$\delta u$       & Change in row offset along the x-axis between rows. \\ \hline
$\delta x$            & Size of a slice along the x-axis.                                                                                 \\ \hline
$\delta y$            & Side of a slice along the y-axis (slit width).                                                                                      \\ \hline
$g_{x,w}$           & Width of gaps between columns along x.                                                                                                                        \\ \hline
$g_{x,d}$           & Depth of gaps (z-coordinate) between columns along x.                                                                                                                        \\ \hline
$g_{y,w}$           & Width of gaps between slices and rows along y.                                                                                                                        \\ \hline
$g_{y,d}$           & Depth of gaps between slices and rows along y.                                                                                                                        \\ \hline
$c_v$               & Curvature of all of the slices$(=1/R)$.                                                                  \\ \hline
$\kappa$             & Conic constant of all of the slices.                                                                                                                                      \\ \hline
\end{tabular}
\label{tab:params}
\end{table}

\section{Glossary of variables} \label{sec:glossary}
A list of variables and their definitions are provided in Table~\ref{tab:glossary}.

\begin{table}[]
\centering
\cprotect\caption{Variables used in this paper. Parameters defined in Tables~\ref{tab:slice-params} and \ref{tab:params} are omitted, as are insignificant variables that are only used once in the context that they are defined.}
\begin{tabular}{|r|p{0.75\linewidth}|}
\hline
Variable       & Description \\ \hline
$c_v$  & The curvature, equal to the inverse of the radius of curvature.. \\ \hline
$\kappa$  & Conic constant. \\ \hline
$\bm{p}=(x_t,y_t,z_t)$  & Starting coordinates of the input ray. $z_t=0$ in global coordinates. \\ \hline
$\bm{\ell}=(l,m,n)$  & Direction cosines of the input ray. \\ \hline
$t$  & Transfer distance of the ray to the surface. \\ \hline
$(x_s,y_s,z_s)$  & Coordinates of the ray incident on the surface. \\ \hline
$(l',m',n')$  & Direction cosines of the refracted (or reflected) ray. \\ \hline
$D,F,G$  & Coefficients of the quadratic equation for $t$. \\ \hline
$\bm{N}=(N_x,N_y,N_z)$  & Surface normal vector. Should be normalized to 1. \\ \hline
$\eta_1$  & Index of refraction of the starting medium. \\ \hline
$\theta_1$  & Angle of incidence. \\ \hline
$\eta_2,\theta_2$  & Index of refraction of the next medium and the angle of refraction.\\ \hline
$\eta_r$  & Ratio of the indices of refraction $\eta_1/\eta_2$. \\ \hline
$\Phi_x,\Phi_y$  & Power of the surface along x and y. Used for the paraxial ray trace. \\ \hline
$l_s,m_s$  & Slopes of the paraxial ray. \\ \hline
$A_{tot}$  & Matrix that encodes the transformation of a slice. \\ \hline
$R_x,R_y$  & Rotation matrices. \\ \hline
$T,T_{OAD}$  & Translation matrices; $T_{OAD}$ is for the off-axis distance for non-planar surfaces. \\ \hline
$x_0,y_0$  & Off-axis distances. \\ \hline
$\Delta X,\Delta Y$  & Dimensions of the image slicer. \\ \hline
$n_{slices}$  & Number of slices in a column of the image slicer. \\ \hline
$n_c,n_r,n_{s,r}$  & Row and column indices. \\ \hline
$n_s$  & Slice index within a column, indexed from zero from the bottom of the image slicer. \\ \hline
$n_{s,r}$  & Slice index within a section, indexed from zero from the bottom of the section. \\ \hline
$(x_c,y_c)$  & Coordinates of the critical point of a slice. \\ \hline
$z_{min},z_{max}$  & The global extrema of the image slicer, which are the mininum and maximum values of the sag. \\ \hline
$u_{min},u_{max}$  & The minimum and maximum values of $u$ for the image slicer, which is the shift of a row along the x-axis. \\ \hline
$v_1,v_2,v_3$  & Placeholder variables for translations defined by the slice parameters (e.g., \cref{eq:v1,eq:v2,eq:v3}. \\ \hline
$A,B,C$  & Coefficients of the quadratic equation for the transformed sag. \\ \hline

\end{tabular}
\label{tab:glossary}
\end{table}

\section{Closed-form solutions} \label{sec:closedform}
In the process of writing this paper, closed-form solutions to the sag, ray transfer distance, and surface normals were discovered for the transformed surfaces. These solutions provide identical results to the process described in Sec.~\ref{sec:implementation.transform} and are marginally more efficient. However, the resulting code is less readable and harder to maintain. The derivation may appear unsightly but is straightforward as it amounts to solving quadratic equations. I verified the solutions using Mathematica and by comparing results to numerically rotated surfaces. See the Github repository for the Mathematica source code.

\subsection{Rotationally symmetric conic}
Obtaining an expression for the transformed sag involves carrying out a matrix multiplication in \cref{eq:transformed-sag} to create a system of equations, which may be solved for $z'(x',y')$. Alternatively, we can pretend that the surface started in the transformed coordinate system and undo the transformation as such:
\begin{equation}
    A_{tot}^{-1} \mathbf{v'} = \mathbf{v}.
\end{equation}
In this case we should set \cref{eq:conic-sag} to $z'$ and solve for $z(x,y)$ instead. We will do this because it is easier to solve. Carrying out the multiplication yields the system of equations
\begin{align}
    x' &= (x+u-s_{yx}) \cos\gamma + (-z+s_{yz}+z_p) \sin\gamma + x_0 + s_{yx}\label{eq:xp}\\
    y' &= y \label{eq:yp} + y_0\\
    z' &= (x-s_{yx}+u) \sin\gamma - (-z+s_{yz}+z_p)\cos\gamma  + s_{yz}\label{eq:zp}\\
    z' &= \frac{c_v (x'^2 + y'^2)}{1 + \sqrt{1 - (1+\kappa)c_v^2(x'^2+y'^2)}}.\label{eq:zp-conic}
\end{align}
We will set the following for brevity:
\begin{align}
    v_1 &= s_{yx}+x_0 \label{eq:v1} \\
    v_2 &= u - s_{yx} \label{eq:v2} \\
    v_3 &= s_{yz} + z_p. \label{eq:v3}
\end{align}
For $\kappa \neq -1$, the transformed sag can be obtained by writing \cref{eq:zp-conic} as
$$
z' c_v (1 + \kappa) = 1 - \sqrt{1 - (1 + \kappa)c_v^2(x'^2 + y'^2)}
$$
and then substituting \cref{eq:xp,eq:yp,eq:zp}. 
Solving for $z$, we obtain
\begin{equation} \label{eq:rot-conic-1}
    z = \begin{cases} 
        \frac{C}{-B+\epsilon\sqrt{B^2-AC}} & \text{if } |\gamma| < \frac{\pi}{2} \\
        \frac{C}{-B-\epsilon\sqrt{B^2-AC}} & \text{if } |\gamma| \geq \frac{\pi}{2}
    \end{cases}
\end{equation}
where
\begin{equation}
    \epsilon = \begin{cases}
    1 & \text{if } \kappa \geq-1\\
    -1 & \text{if } \kappa < -1
    \end{cases}
\end{equation}
and
\begin{align}
    \begin{split}\label{eq:rot-conic-A}
        A ={}& c_v (1+\kappa)(2+\kappa+\kappa\cos 2\gamma)
    \end{split}\\
    \begin{split} \label{eq:rot-conic-B}
        B ={}& -(1+\kappa) \Big(
            -2(-1 + c_v(1+\kappa)s_{yz})\cos\gamma \\
            &+ c_v\left((2+\kappa)v_3 + \kappa v_3 \cos 2\gamma + 2 v_1 \sin\gamma - \kappa (v_2 + x) \sin 2\gamma \right)
        \Big)
    \end{split}\\
    \begin{split} \label{eq:rot-conic-C}
        C ={}& (1+\kappa)\Big(
            -4 s_{yz}
            + 2 c_v ((1 + \kappa) s_{yz}^2 + v_1^2 + (y + y_0)^2) \\
            &+ c_v (2 + \kappa)(v_2^2 + v_3^2 + 2 v_2 x + x^2) \\
            &+ 4 (v_3 - c_v (1 + \kappa) s_{yz} v_3 + c_v v_1 (v_2 + x)) \cos\gamma  \\
            &- c_v \kappa (v_2 - v_3 + x)(v_2 + v_3 + x) \cos 2\gamma \\
            &+ 4 \left( (-1 + c_v (1 + \kappa) s_{yz}) v_2 + c_v v_1 v_3 - x + c_v (1 + \kappa) s_{yz} x \right)  \sin\gamma\\
            &- 2 c_v \kappa v_3 (v_2 + x) \sin 2\gamma.
        \Big)
    \end{split}
\end{align}
The case of $\kappa=-1$ can be dealt with separately. \Cref{eq:zp-conic} reduces to
$$
    z' = \frac{c_v (x'^2 + y'^2)}{2}.
$$
The solution for a parabola also takes the form of a quadratic that satisfies \cref{eq:rot-conic-1}, instead with coefficients
\begin{align}
    \begin{split} \label{eq:par-A}
        A_p ={}& c_v \sin^2 \gamma
    \end{split} \\
    \begin{split} \label{eq:par-B}
        B_p ={}& -\Big(
            \cos \gamma
            + c_v (v_2 + x) \cos \gamma \sin \gamma
            + c_v \sin \gamma (v_1 + v_3 \sin \gamma)
        \Big)
    \end{split} \\
    \begin{split} \label{eq:par-C}
        C_p ={}& -2 s_{yz}
        + c_v \left(v_1^2 + (y + y_0)^2\right) \\
        &+ 2 (v_3 + c_v v_1 (v_2 + x)) \cos \gamma
        + c_v (v_2 + x)^2 \cos^2 \gamma \\
        &- 2 (v_2 - c_v v_1 v_3 + x) \sin \gamma
        + c_v v_3^2 \sin^2 \gamma
        + c_v v_3 (v_2 + x) \sin 2\gamma.
    \end{split}
\end{align}

In general, rotating a function does not guarantee that it will be a function post-rotation. Consider the extreme case of rotating by \qty{90}{\degree} where a given $(x,y)$ corresponds to either multiple $z$ values or is undefined. These undefined regions are indicated by $B^2 < 4AC$. We can set $z$ to zero in these cases to prevent errors at runtime. Choosing the positive or negative solution of the quadratic is somewhat arbitrary near $|\gamma| = \frac{\pi}{2}$ but matters for less extreme values of $\gamma$, which comprise the vast majority of cases.

The transfer distance for a single slice is given by plugging \cref{eq:transfer-1,eq:transfer-2,eq:transfer-3} into \cref{eq:rot-conic-1} and solving for $t$. Because the sag is now in global coordinates, $z_t$ should be set to zero. The result is
\begin{equation} \label{eq:trans-dist-1}
    t = \begin{cases} 
        \frac{G}{-F+\sqrt{F^2-DG}} & \text{if } |\gamma| < \frac{\pi}{2} \\
        \frac{G}{-F-\sqrt{F^2-DG}} & \text{if } |\gamma| \geq \frac{\pi}{2}
    \end{cases}
\end{equation}
where
\begin{align}
    \begin{split} \label{eq:trans-dist-D}
        D ={}& 2 c_v \kappa (n \cos\gamma + l \sin\gamma)^2 + 2 c_v
    \end{split} \\
    \begin{split} \label{eq:trans-dist-F}
        F ={}& c_v \left( - (2 + \kappa) n v_3 + (2 + \kappa) l (v_2 + x_t) + 2 m (y_0 + y_t) \right) \\
        &+ 2 \left( n (-1 + c_v (1 + \kappa) s_{yz}) + c_v l v_1 \right) \cos\gamma \\
        &- c_v \kappa (n v_3 + l (v_2 + x_t)) \cos 2\gamma \\
        &+ 2 \left( l (-1 + c_v (1 + \kappa) s_{yz}) - c_v n v_1 \right) \sin\gamma \\
        &+ c_v \kappa (-l v_3 + n (v_2 + x_t)) \sin 2\gamma
    \end{split} \\
    \begin{split} \label{eq:trans-dist-G}
        G ={}& -4 s_{yz} + 2 c_v \left( (1 + \kappa) s_{yz}^2 + v_1^2 + (y_0 + y_t)^2 \right) \\
        &+ c_v (2 + \kappa)(v_2^2 + v_3^2 + 2 v_2 x_t + x_t^2) \\
        &+ 4 \left( v_3 - c_v (1 + \kappa) s_{yz} v_3 + c_v v_1 (v_2 + x_t) \right) \cos\gamma \\
        &- c_v \kappa (v_2 - v_3 + x_t)(v_2 + v_3 + x_t) \cos 2\gamma \\
        &+ 4 \left( (-1 + c_v (1 + \kappa) s_{yz}) v_2 + c_v v_1 v_3 - x_t + c_v (1 + \kappa) s_{yz} x_t \right) \sin\gamma \\
        &- 2 c_v \kappa v_3 (v_2 + x_t) \sin 2\gamma.
    \end{split}
\end{align}
For $\kappa=-1$, we instead use \cref{eq:par-A,eq:par-B,eq:par-C} and obtain
\begin{align}
    \begin{split} \label{eq:par-D}
        D_p ={}& c_v \left( m^2 + \left( l \cos\gamma - n \sin\gamma \right)^2 \right)
    \end{split} \\
    \begin{split} \label{eq:par-F}
        F_p ={}& c_v m (y_0 + y_t)
        + c_v l (v_2 + x_t) \cos^2\gamma \\
        &- \sin\gamma \left( l + c_v n v_1 + c_v n v_3 \sin\gamma \right) \\
        &- \cos\gamma \left( n - c_v l v_1 + c_v (-l v_3 + n (v_2 + x_t)) \sin\gamma \right)
    \end{split} \\
    \begin{split} \label{eq:par-G}
        G_p ={}& -2 s_{yz}
        + c_v \left( v_1^2 + (y_0 + y_t)^2 \right) \\
        &+ 2 (v_3 + c_v v_1 (v_2 + x_t)) \cos\gamma
        + c_v (v_2 + x_t)^2 \cos^2\gamma \\
        &- 2 (v_2 - c_v v_1 v_3 + x_t) \sin\gamma
        + c_v v_3^2 \sin^2\gamma
        + c_v v_3 (v_2 + x_t) \sin 2\gamma.
    \end{split}
\end{align}
The solutions are valid for all values of $\gamma$ and assume that the direction cosines are normalized such that $l^2 + m^2 + n^2 = 1.$ The ray tracing algorithm described in Sec.~\ref{sec:implementation.ray} is still required because the slice parameters are dependent on the point of intersection of the ray. When applied across the entire image slicer, \cref{eq:trans-dist-1} becomes a discontinuous transcendental equation with no closed-form solution.

The surface normal vector can be found similarly to \cref{eq:sag-surface-for-grad} except using \cref{eq:rot-conic-1} for the sag. Let $\omega=-B \pm \sqrt{B^2-AC}$. Applying the chain rule,
\begin{equation}
    \frac{\partial \omega}{\partial x} = -\frac{\partial B}{\partial x} \pm \frac{1}{2\sqrt{B^2-AC}}\Big( 2B \frac{\partial B}{\partial x} - C \frac{\partial A}{\partial x}-A \frac{\partial C}{\partial x}\Big)
\end{equation}
\begin{equation}\label{eq:dervx}
    \frac{\partial}{\partial x} \Bigg( \frac{C}{-B\pm\sqrt{B^2-AC}} \Bigg)= \frac{1}{\omega^2} \Big( \omega \frac{\partial C}{\partial x} - C \frac{\partial \omega}{\partial x} \Big).
\end{equation}
The partial derivative about y is found the same way. The sign differences $(\pm)$ correspond to the positive and negative quadratic solutions of \cref{eq:rot-conic-1}, respectively. The partial derivatives of \cref{eq:rot-conic-A,eq:rot-conic-B,eq:rot-conic-C} are
\begin{align}
    \begin{split}\label{eq:dervAx}
        \frac{\partial A}{\partial x} ={}& 0
    \end{split}\\
    \begin{split}\label{eq:dervBx}
        \frac{\partial B}{\partial x} ={}& -c_v \kappa (1+\kappa)\sin 2\gamma
    \end{split}\\
    \begin{split}\label{eq:dervCx}
        \frac{\partial C}{\partial x} ={}& (1 + \kappa)\Big(
            2 c_v(2 + \kappa)(v_2 + x)
            + 4 c_v v_1 \cos\gamma \\
            &\quad - c_v \kappa (v_2 - v_3 + x) \cos 2\gamma
            - c_v \kappa (v_2 + v_3 + x) \cos 2\gamma \\
            &\quad - 4(-1 + c_v(1 + \kappa) s_{yz}) \sin\gamma
            + 2 c_v \kappa v_3 \sin 2\gamma
        \Big)
    \end{split}\\
    \frac{\partial A}{\partial y} = {}& 0\\
    \frac{\partial B}{\partial y} = {}& 0\\
    \frac{\partial C}{\partial y} = {}& 4c_v (1+\kappa)(y+y_0). \label{eq:dervyC}
\end{align}
For $\kappa=-1$,
\begin{align}
    \frac{\partial B_p}{\partial x} &= -c_v \cos\gamma \sin\gamma\\
    \frac{\partial C_p}{\partial x} &= -2 \sin\gamma + 2 c_v \cos\gamma \left( v_1 + (v_2 + x) \cos\gamma + v_3 \sin\gamma \right)
\\
    \frac{\partial C_p}{\partial y} &= 2c_v(y+y_0)
\end{align}
with the remaining partial derivatives being equal to zero. The surface normal vector is then
\begin{equation}\label{eq:dervx}
    \bm{N} = \frac{1}{\omega^2} \Big( \omega \frac{\partial C}{\partial x} - C \frac{\partial \omega}{\partial x} \Big) \bm{\hat{x}}+ \frac{1}{\omega^2} \Big( \omega \frac{\partial C}{\partial y} - C \frac{\partial \omega}{\partial y} \Big) \bm{\hat{y}} - \bm{\hat{z}}.
\end{equation}

\subsection{Flat surfaces}
Following the same process as above but using \cref{eq:flat-sag-base} for the sag and \cref{eq:flat-transformation} for the transformation, we obtain the the system of equations
\begin{align}
    \begin{split}\label{eq:flat-1}
        x' ={}& (x-s_{yx}+u) \cos(\beta+\gamma) + (-z+s_{yz}+z_p) \sin(\beta+\gamma) + s_{yx}
    \end{split}\\
    \begin{split}\label{eq:flat-2}
        y' ={}& (y-s_{xy})\cos\alpha-(s_{xz}-s_{yz}+(-z+s_{yz}+z_p)\cos(\beta+\gamma) \\
        &-(x-s_{yx}+u)\sin(\beta+\gamma)) \sin\alpha + s_{xy}
    \end{split}\\
    \begin{split}\label{eq:flat-3}
        z' ={}& (s_{xy}-y)\sin\alpha - (s_{xz}-s_{yz} + (-z + s_{yz} + z_p) \cos(\beta+\gamma) \\
        &-(x-s_{yx}+u)\sin(\beta+\gamma))\cos\alpha + s_{xz}
    \end{split}\\
    \begin{split}\label{eq:flat-4}
        z' ={}& 0.
    \end{split}
\end{align}
We only need \cref{eq:flat-3,eq:flat-4} to obtain
\begin{equation}
    z = \sec(\beta+\gamma) (s_{xz}-s_{yz}-s_{xz}\sec\alpha + (y-s_{xy}) \tan\alpha) - (x-s_{yx}+u) \tan(\beta+\gamma) + s_{yz} + z_p
\end{equation}
and solving for the ray transfer distance in the same manner as before,
\begin{multline}
    t = \frac{1}{n-m\sec(\beta+\gamma)\tan\alpha+l\tan(\beta+\gamma)} \Big(
    \sec(\beta+\gamma)(s_{xz}-s_{yz}-s_{xz}\sec\alpha+(y_t-s_{xy})\tan\alpha)\\
    - (x_t-s_{yx}+u)\tan(\beta+\gamma) + s_{yz}+z_p
    \Big).
\end{multline}
The global extrema can be computed without solving for critical points because a plane has no vertex. The surface normal is
\begin{equation}
    \mathbf{N} = -\tan(\beta+\gamma) \mathbf{\hat{x}} + \sec(\beta+\gamma)\tan\alpha \mathbf{\hat{y}}
\end{equation}

These expressions are undefined when the plane is rotated by $\pm90^\circ$ such that it is rotated perpendicular to the line of sight, which is when $\cos(\beta+\gamma)=0$ or $\cos\alpha=0$. This is unphysical and should not occur if the DLL is being used as intended. We can place a lower limit on these values to prevent a division by zero error.

\subsection{Cylindrical surfaces}
Is is possible to follow the process yet again with \cref{eq:sag-cylinder} for the sag and \cref{eq:cylinder-transformation} for the transformation. Doing so yields the following system of equations:
\begin{align}
    \begin{split}\label{eq:1}
        x' ={}& (x-s_{yx}+u)\cos\gamma + (-z+s_{yz}+z_p)\sin\gamma + s_{yx} + x_0
    \end{split}\\
    \begin{split}\label{eq:2}
        y' ={}& (y-s_{xy})\cos\alpha-(s_{xz}-s_{yz}+(-z+s_{yz}+z_p)\cos\gamma \\
        & - (x-s_{yx}+u)\sin\gamma) \sin\alpha + s_{xy}
    \end{split}\\
    \begin{split}
        z' ={}& (s_{xy}-y)\sin\alpha - (s_{xz}-s_{yz} + (-z + s_{yz} + z_p) \cos\gamma \\
        &-(x-s_{yx}+u)\sin\gamma)\cos\alpha + s_{xz}
    \end{split}\\
    \begin{split}
        z' ={}& \frac{c_v x'^2}{1 + \sqrt{1 - (1+\kappa)c_v^2x'^2}}.
    \end{split}
\end{align}
This is quite similar to \cref{eq:flat-1,eq:flat-2,eq:flat-3} except with an additional translation $x_0$ and a more complex expression for the untransformed sag. The resulting solutions are not provided here as they involve a very large number of terms. However, it is still possible to reduce the expressions into quadratic coefficients was done for the other two surface types. The reader is deferred to the Mathematica notebook in the Github repository if they would like to see the full solutions for cylinders.

\subsection*{Disclosures}
The author declares that there are no financial interests, commercial affiliations, or other potential conflicts of interest that could have influenced the objectivity of this research or the writing of this paper.

\subsection* {Code, Data, and Materials Availability}
The majority of the code developed for this project is available in a Github repository, which is currently hosted at the following link: \url{https://github.com/ellenle3/ifugen}. Code provided by Zemax cannot be included with the repository, including the \verb|usersurf.h| header file that is necessary to compile the DLLs. This file along with example DLLs included with Zemax can typically be found in \verb|C:\\...\\Documents\\Zemax\\DLL\\|. At the time of writing, compiled versions of the DLLs are not available in the repository due to a few known issues that may cause Zemax to crash, mainly instantiating multiple instances of the same DLL or running POP with a large spatial sampling. The author is actively working to resolve these issues and improve documentation. However, the readers are welcome to compile the DLLs and try them at their own risk.

\subsection* {Acknowledgments}
The author is funded by the Infrared Telescope Facility which is operated by the University of Hawaii under contract 80HQTR19D0030 with the National Aeronautics and Space Administration. The author would like to thank Michael Connelley for proofreading this paper, sharing the design of SPECTRE, and being supportive of this project despite the fact that it is wholly unrelated to her dissertation. She would also like to thank Renate Kupke, Haosheng Lin, and Deno Stelter for discussions regarding their process of designing image slicer IFUs. The author was strongly motivated by the challenges that her colleagues have faced and feels that the knowledge they have imparted on her is invaluable.


\bibliography{report}   
\bibliographystyle{spiejour}   


\vspace{2ex}\noindent\textbf{Ellen Lee} is a PhD candidate at the University of Hawai'i's Institute for Astronomy. Her primary area of research is in the facility integration of adaptive secondary mirrors to enhance instrument performance via wide-field adaptive optics.

\listoffigures
\listoftables

\end{document}